\documentclass[
reprint, 
superscriptaddress,
prx]{revtex4-2}
\usepackage{amsmath, amssymb}
\usepackage{graphicx} 
\usepackage{upgreek} 
\usepackage{natbib} 
\usepackage{fancyhdr}
\usepackage{multirow}
\usepackage{bm}
\usepackage{array}
\usepackage[linkcolor=magenta, citecolor=magenta, colorlinks=True]{hyperref}

\usepackage{color}
\usepackage{xcolor}

\newcommand{\blue}[1]{\textcolor{blue}{#1}}

\begin{document}
\title{Signal crosstalk in a flip-chip quantum processor}

\def\CHALMERS{Chalmers University of Technology, Gothenburg 412 96, Sweden}
\def\VTT{VTT Technical Research Centre of Finland, VTT Espoo FI-02044 , Finland}
\author{Sandoko Kosen}
\email{sandoko.kosen@gmail.com}
\affiliation{\CHALMERS}

\author{Hang-Xi Li}
\affiliation{\CHALMERS}

\author{Marcus Rommel}
\affiliation{\CHALMERS}

\author{Robert Rehammar}
\affiliation{\CHALMERS}

\author{Marco Caputo}
\affiliation{\VTT}

\author{Leif Gr{\"o}nberg}
\affiliation{\VTT}

\author{Jorge Fern{\'a}ndez-Pend{\'a}s}
\affiliation{\CHALMERS}

\author{Anton Frisk Kockum}
\affiliation{\CHALMERS}

\author{Janka Bizn{\'a}rov{\'a}}
\affiliation{\CHALMERS}

\author{Liangyu Chen}
\affiliation{\CHALMERS}

\author{Christian Kri\v{z}an}
\affiliation{\CHALMERS}

\author{Andreas Nylander}
\affiliation{\CHALMERS}

\author{Amr Osman}
\affiliation{\CHALMERS}

\author{Anita Fadavi Roudsari}
\affiliation{\CHALMERS}

\author{Daryoush Shiri}
\affiliation{\CHALMERS}

\author{Giovanna Tancredi}
\affiliation{\CHALMERS}

\author{Joonas Govenius}
\affiliation{\VTT}

\author{Jonas Bylander}
\email{jonas.bylander@chalmers.se}
\affiliation{\CHALMERS}

\date{\today}

\begin{abstract}
Quantum processors require a signal-delivery architecture with high addressability (low crosstalk) to ensure high performance already at the scale of dozens of qubits. 
Signal crosstalk causes inadvertent driving of quantum gates, which will adversely affect quantum-gate fidelities in scaled-up devices. 
Here, we demonstrate packaged flip-chip superconducting quantum processors with signal-crosstalk performance competitive with those reported in other platforms.
For capacitively coupled qubit-drive lines, we find on-resonant crosstalk better than $-27\,$dB (average $-37\,$dB).
For inductively coupled magnetic-flux-drive lines, we find less than 0.13\,\% direct-current flux crosstalk (average 0.05\,\%). These observed crosstalk levels are adequately small and indicate a decreasing trend with increasing distance, which is promising for further scaling up to larger numbers of qubits. 
We discuss the implication of our results for the design of a low-crosstalk, on-chip signal delivery architecture, including the influence of a shielding tunnel structure, potential sources of crosstalk, and estimation of crosstalk-induced qubit-gate error in scaled-up quantum processors. 
\end{abstract}

\maketitle
\section{Introduction}
The quest for demonstrating quantum computational advantage and fault-tolerant quantum computation has inspired the realisation of integrated quantum processors~\cite{googlequantumai2021, kim2023, ryan-anderson2021, bluvstein2023}. 
A prominent physical platform is based on superconducting qubits, which are typically sub-millimeter in size, made of lithographically defined thin-film devices on low-loss substrates, and operate at frequencies below 10\,GHz in a cryogenic environment. Scaling up superconducting quantum processors requires microchip integration in an extensible design while maintaining high-fidelity qubit performance through predictable device parameters and high-yield fabrication. 

\begin{figure}[t]
    \includegraphics{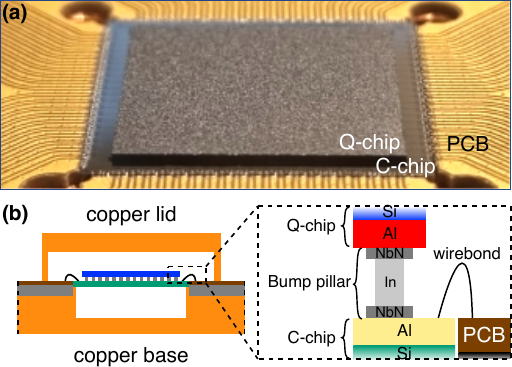}
    \caption{A packaged 25-qubit superconducting quantum processor. (a) Close-up photo of the flip-chip processor packaged with wirebonds and printed-circuit board (PCB) technology. (b) Illustration of the cross section showing the copper lid and base, with a zoomed-in illustration showing the material stack (not to scale; Si: silicon, Al: aluminium, NbN: niobium nitride, In: indium).}\label{fig1}
\end{figure}

An outstanding problem for extensible designs is on-chip signal routing that enables components within the processor to be selectively addressed and read out. 
In current technology, signals are usually routed from the edge of the chip, where wirebonds make connection to a printed circuit board and further on to connectors on a microwave package. 
If the signal is insufficiently shielded, crosstalk ensues and results in quantum-gate errors that render quantum computation infeasible.
Active crosstalk suppression techniques, which involve characterising the crosstalk matrix and applying its inverse~\cite{dai2021, sung2021, nuerbolati2022, barrett2023, zhang2023, dai2024}, are prohibitively challenging at scale. We must thus ensure that the circuit architecture supports adequate passive crosstalk suppression by proper routing and shielding techniques.  
Therefore, the roadmap for next-generation quantum processors requires an understanding of the influence of densely routed signal lines and an informed strategy for signal delivery with low crosstalk. 
This challenge has not received the attention it deserves.

Signal crosstalk can be investigated in dedicated devices, yielding data that is conceptually simple to interpret. Alternatively, it can be performed with devices that are designed for actual implementations of quantum algorithms. Such an approach enables a more accurate assessment of the relevant crosstalk level and can tease out potential crosstalk mechanisms that are dominant at a larger scale, although at the expense of a more complicated data interpretation. This work adopts the second approach, but we consider the two approaches to be equally valuable, necessary, and complementary of each other in the long run.

In this work, we demonstrate packaged multi-qubit processors consisting of 25 qubits in a flip-chip module, as shown in fig.\,\ref{fig1}. 
The processor is designed according to a repeating signal-line routing pattern, which is, in principle, scalable to hundreds of qubits.
We characterize the microwave crosstalk (selectivity of driving) of qubit drive lines (``xy-lines''), which are capacitively coupled to their target qubits, and of flux lines (``z-lines''), which are inductively coupled to frequency-tunable couplers that mediate two-qubit gates.

The results show crosstalk performance approaching average values of $-40$\,dB for microwave-drive xy crosstalk and 0.05\,\% for direct-current flux crosstalk, competitive with reported performance from other superconducting \cite{krinner2022, spring2022, dai2021, zhang2023, barrett2023} and non-superconducting \cite{wang2020, xia2015, lawrie2023} platforms. 
We show that a substantial contribution to the flux crosstalk is due to the proximity of the grounded end of a victim z-line to a neighbouring source z-line, and it can be suppressed either by a shielding tunnel structure or by adequate separation. On the other hand, enclosing the xy-lines with shielding tunnels does not improve xy crosstalk despite the denser routing layout, suggesting that the intrinsic crosstalk level due to direct capacitive interaction is better than the observed performance (simulation of a simplified model of our processor also suggests that it is the case).

\begin{figure*}[t!]
    \includegraphics{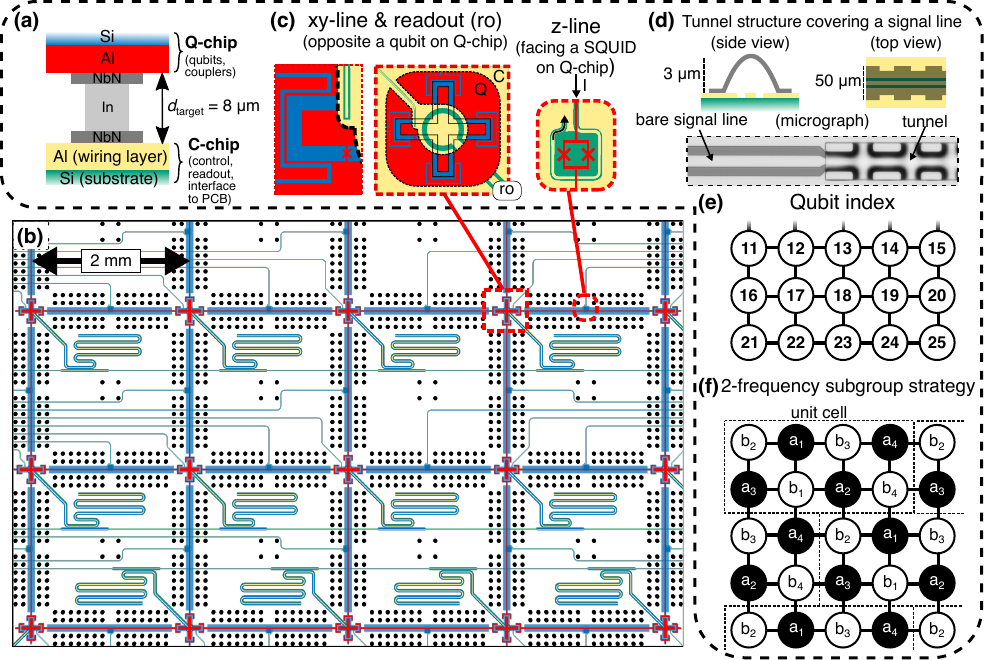}
    \caption{Quantum processor architecture. 
    (a) Two-tier flip-chip material stack (not to scale). 
    (b) Overlapped view of Q-chip (qubits and couplers) and  C-chip (control and readout elements) for the bottom 15 qubits of a 25-qubit QPU, showing the bump pattern (black dots). For an easier view, both ground planes are removed from this illustration. 
    (c, left) A transmon (Xmon) qubit located on the Q-chip, with xy-line (qubit drive) and readout resonator located on the C-chip. The cutout on the Q-chip layer shows the end parts of the xy-line and of the resonator. The panel on the left shows a zoomed-in picture of the region around the tip of the xy-line.
    (c, right) A z-loop (flux loop) on the C-chip whose center lies directly opposite a SQUID on the Q-chip (wiring and substrate on the Q-chip not shown). 
    (d) Aluminium tunnel structure shielding a signal line. The micrograph shows one part of the transmission line covered with a tunnel. 
    (e) Qubit indexing used in this paper. 
    (f) Two-frequency subgroup strategy: the filled and empty circles indicate transmons with two different anharmonicities; a$_k$ and b$_k$ indicate different qubit frequencies (a$_k$ $<$ b$_k$), giving a total of eight different frequencies that are tiled according to the schematic.}\label{fig2}
\end{figure*}

It is generally desirable for the crosstalk level to be not only low but also to decrease with increasing separation, and our results suggest such a trend. To quantify, a reasonable xy-crosstalk magnitude to aim for is such that the total single-qubit gate error does not exceed the $0.1\,\%$ threshold recommended for quantum-error correction \cite{martinis2015}. We provide numerical estimates of the total single-qubit gate error based on the measured xy-crosstalk and sketch the further improvement required for the total error to stay below the $0.1\,\%$ threshold for processors at the 100-qubit level.
We discuss potential sources of crosstalk, its impact on gate fidelities, and its implication for future designs of low-crosstalk processors.

\section{Quantum processor architecture\label{main_quantumprocessorarchitecture}}
Our superconducting quantum processing unit (QPU) consists of a two-tiered architecture \cite{rosenberg2017, foxen2018, gold2021, kosen2022} separating the circuit into a qubit chip (``Q-chip") and a control chip (``C-chip"), see figs.\,\ref{fig1} and \ref{fig2}(a). 
The Q-chip comprises 25 fixed-frequency transmon (Xmon) qubits~\cite{Koch2007,Schreier2008,Barends2013} and 40 frequency-tunable two-qubit couplers laid out on a square grid with 2\,mm pitch \cite{mckay2016}. This relatively large pitch is a choice out of convenience, since the QPU performance is limited by gate performance rather than the number of devices that can be fitted on a die. 
The C-chip comprises a signal delivery system routed through coplanar waveguide transmission lines on the chip surface. This signal-routing strategy is extensible to hundreds of qubits.
Figure \ref{fig2}(b) shows a part of the flip-chip module with the Q-chip overlaid on the C-chip. 

Figure \ref{fig2}(c) shows the control wires' coupling points to the qubit and coupler. 
The qubit-readout resonators [ro in fig.\,\ref{fig2}(c)] are quarter-wavelength resonators with the open end positioned directly opposite from a qubit. 
The resonators are undulated to reduce coupling with neighbouring couplers.  
The open ends of the qubit-control lines (xy) are positioned to ensure adequate capacitive coupling while minimizing Purcell decay of the qubit into its own xy-line. 
Finally, the coupler-control lines (z) each terminate in a loop of wire that is shorted to ground at one end without a ground plane within the loop. 
The current in the z-loop couples magnetic flux into the SQUID of the coupler positioned opposite the loop; this SQUID connects the coupler's island to the ground plane.

The second processor discussed in this work has aluminium tunnel structures~\cite{chen2014, dunsworth2018} added to the C-chip, shielding the signal lines as shown in fig.\,\ref{fig2}(d). 
These tunnels extend across the signal lines and connect the ground planes on either side.

The Q-chip and C-chip consist of 12\,mm$\times$12\,mm and 14.3\,mm$\times$14.3\,mm silicon dies, respectively, which were aligned and joined together by flip-chip bonding with target inter-chip separation of $8\,\upmu$m. 
The bump-bond layout is symmetrical across the dies with 2900 superconducting bumps connecting the ground planes of the two tiers, with denser distribution surrounding the qubit and coupler devices. 
The bumps consist of evaporated indium cylinders (25\,$\upmu$m wide pre-compression) with a niobium-nitride underbump metallisation layer \cite{kosen2022}. 

The two-tier flip-chip module represents a straightforward method for 3D integration of QPUs, offering more flexibility for signal-wire routing and qubit-array layouts than planar layouts, with a modest increase of fabrication process complexity.
It also enables the use of different fabrication processes for the two chips.
In more advanced 3D-integration implementations, superconducting bumps can be used to pass signals between tiers and to provide shielding between components; 
hard stops (e.g. posts or pillars) can help achieve precise inter-chip separation \cite{niedzielski2019, norris2024}; 
furthermore, substrates with metallised through-silicon vias (TSVs) and buried conductors can help create signal redistribution layers \cite{vahidpour2017, mallek2021a, grigoras2022}. 

Aluminium wirebonds along the periphery of the C-chip transfer signals to and from a multilayer printed circuit board, see fig.\,\ref{fig1}. Two wirebonds are placed on every signal launchpad, and two grounding wirebonds in-between neighbouring signal lines. 
This wiring method works up to hundreds of wires, but begins to scale poorly beyond that, which is why ultimately higher-density interconnects and packaging solutions, e.g. ball-grid arrays or land-grid arrays, need to be adapted to superconducting QPU technology. 

Our flip-chip fabrication process has been demonstrated to be compatible with qubit (transmon) coherence at the 100-$\upmu$s level~\cite{kosen2022}. The wiring layer (signal and ground) on each chip is fabricated in an aluminium process, and the Josephson junctions consist of an angle-evaporated Al/AlOx/Al sandwich using the two-step Manhattan process~\cite{burnett2019, osman2021}. 
The fabrication and packaging steps are outlined in Appendix \ref{app_fabricationpackaging}. Details on the flip-chip bonding processes can be found in the supplementary material of Ref.\,\cite{kosen2022}.

The qubits are labelled q$_i$, where the index $i$ starts in the top left corner of the chip; see fig.\,\ref{fig2}(e), which shows q$_{11}$ to q$_{25}$.
The element xy$_i$ is the corresponding xy-line for q$_i$. 
The couplers and the corresponding z-lines are labelled cp$_i$ and z$_i$, where the index $i$ is a pair of indices indicating the two qubits that are connected by the coupler; e.g. cp$_{(11,16)}$ is the coupler connecting q$_{11}$ and q$_{16}$.
The closest separation between a qubit and another xy-line (excluding its own xy-line) is about $500\,\upmu$m.

The routing of signal lines is designed to be as identical as possible for these rows of qubits: \@ q$_6$ to q$_{10}$, q$_{11}$ to q$_{15}$, and q$_{16}$ to q$_{20}$. 
The control elements (xy, z) are always routed to the desired positions from the corridor above the qubits, while the readout elements are positioned below; see fig.\,\ref{fig2}(b). 
The signal lines are routed horizontally to the periphery of the chip. In each corridor, there is always a z-line in-between the closest pair of xy-lines. 
These z-lines are needed to control horizontally oriented tunable couplers. 
The vertically oriented couplers, however, are controlled by z-lines that are routed side-by-side. 
Nearest-neighbour control lines are separated by at least $100\,\upmu$m. 
For readout purposes, a feedthrough transmission line is routed across each corridor and is coupled to five $\lambda/4$-resonators. 
Once a signal lines emerge at the qubit-array perimeter, it is routed to the closest available wirebond launchpad. 
For qubits in the top row (q$_1$ to q$_5$), bottom row (q$_{21}$ to q$_{25}$), and the corresponding horizontally oriented couplers, the control lines are routed vertically towards the closest wirebond launchpads. 
No bumps or airbridge crossovers are used for signal delivery, i.e. there is no intersection of signal lines.

The allocation of qubit frequency and anharmonicity follows the two-frequency subgroup strategy designed to be compatible with implementation of parametric CZs and iSWAPs that avoids frequency collision between neighbouring couplers and minimizes crosstalk due to frequency crowding \cite{osman2023}. Each subgroup, labelled $\{\mathrm{a}_k\}$ and $\{\mathrm{b}_k\}$ ($k=1,2,3,4$), has four qubit frequencies that are distributed around a central frequency; subgroup $\mathrm{a}$ has lower frequencies and anharmonicities than subgroup $\mathrm{b}$.
The layout has a unit cell of $2\times4$ qubits that can be repeatedly tiled as shown in fig.\,\ref{fig2}(f). An alternative visualisation is the following: a qubit with frequency in one subgroup is pairwise coupled to qubits with frequencies in the other subgroup.

The readout-resonator frequency allocation is set up to ensure that, at the targeted inter-chip separation, the qubit-resonator frequency detuning is generally between 2.1 and 2.6\,GHz, which provides sufficient leeway for tolerating frequency shift in case the achieved chip separation is off-target. 
The coupler frequency at zero flux bias is designed to be above the readout-resonator frequency. 

We designed and simulated the flip-chip QPU layout using a combination of the IBM Qiskit Metal design toolkit~\cite{shah2023}, ANSYS electromagnetic simulation software~\cite{zotero-4428}, and L-Edit layout editor~\cite{zotero-4429}. The simulation technique for individual qubits and resonators is described in Ref.\,\cite{kosen2022}.

Refer to Appendix~\ref{app_targetfrequencies} for the chosen numerical values of qubit frequencies, anharmonicities, readout resonator frequencies, and coupler frequencies at zero flux bias. Appendix~\ref{app_measuredparameters} lists the measured frequency parameters and coherence performances of both processors.
\section{Aggregate crosstalk performance}

\begin{figure}[t]
\centering
  \includegraphics{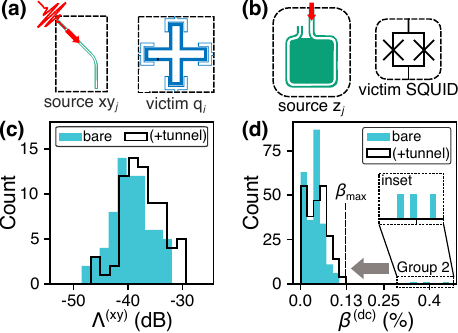}
  \caption{Aggregate performance of xy crosstalk $\Lambda^\mathrm{(xy)}$, and dc-flux crosstalk $\beta^\mathrm{(dc)}$. (a) Illustration of xy crosstalk from a source xy-line ($j$) to a victim qubit ($i$), and (b) dc-flux crosstalk from a source z-line ($j$) to a victim coupler's SQUID loop ($i$).
  (c) Histogram of the on-resonant xy crosstalk, $\Lambda_{i,j}$, measured on 72 pairs for one flip-chip module with bare transmission lines (labelled ``bare"), 
  and another module with a majority of the transmission lines covered by ``tunnel"structures (labelled ``+tunnel"). 
  (d) Histogram of the dc-flux crosstalk, $\beta^{\mathrm{(dc)}}$, measured on 240 pairs. Data indicated as Group 2 are associated with a subset of pairs whose z-lines are nearestneighbour. With the tunnels, the maximum value of dc-flux crosstalk drops to $\beta_{\mathrm{max}}=0.13\,\%$.}\label{fig3}
\end{figure}

Here we quantify the crosstalk (unintended driving) affecting qubits and couplers due to signals applied to the circuit's various xy-lines and z-lines, respectively.
To begin with the microwave crosstalk of xy-lines, consider any pair of a victim qubit (q$_i$) and a source xy-line (xy$_{j\ne i}$) as shown in fig.\,\ref{fig3}(a). 
The crosstalk is quantified by the rotation of the quantum state vector of q$_i$ due to a signal being delivered through xy$_j$ aiming to control its corresponding qubit q$_j$. 
In practice, we compare the Rabi frequency of the victim qubit, $\Omega_{i,j}$, with that of the qubit of the source xy-line, $\Omega_{j,j}$, for the same xy$_j$ signal amplitude. Due to the low-crosstalk nature, we always drive the victim qubit on resonance.
Their ratio, expressed in decibel (dB), is the on-resonant xy crosstalk $\Lambda_{i,j}$ (referred to as xy selectivity in \cite{spring2022}),
\begin{equation} \label{eq:xy_crosstalk_definition}
    \Lambda_{i,j} = 10\times\mbox{log}_{10}\left(\frac{\Omega_{i,j}}{\Omega_{j,j}}\right)^2.
\end{equation}
The lower the magnitude of $\Lambda_{i,j}$, the better that xy-line is at selectively driving its corresponding qubit compared to any other qubit (lower crosstalk).

We also characterise the magnetic-flux crosstalk in the z-lines by examining changes in the magnetic flux ($\Delta\Phi_i$) that threads the SQUID of a victim coupler cp$_i$ due to direct-current (dc) or alternating-current (ac) signals on a source z-line (z$_{j\ne i}$), aimed to control its corresponding coupler cp$_j$, see fig.\,\ref{fig3}(b). The magnetic-flux crosstalk is defined as \cite{abrams2019}
\begin{equation} \label{eq:z_crosstalk_definition}
    \beta_{i,j} = \left|\frac{\Delta\Phi_i}{\Delta\Phi_j}\right| . 
\end{equation}
Generally, $\Phi_i= \Phi^\mathrm{(dc)}_i+\Phi^\mathrm{(ac)}_i\cos{(\omega^\mathrm{(ac)}_i t+\theta_i)}$, where $\omega^\mathrm{(ac)}_i$ is the angular frequency of the ac-flux drive signal and $\theta_i$ is its phase offset. In practice, the dc-flux and the ac-flux crosstalk, i.e. $\beta_{i,j}^\mathrm{(dc)}$ and  $\beta_{i,j}^\mathrm{(ac)}$, are characterised separately.

The measurement procedures for the xy, dc-flux, and ac-flux crosstalk are discussed in Appendix~\ref{app_meastechnique}. As there are also some defective elements in these first-generation processors, the datasets presented in this section are only associated with source-victim pairs that are functional in both processors (bare waveguides and with tunnels) to allow fair comparisons. The complete datasets can be found in Appendix \ref{app_completedatasets}.

Figure~\ref{fig3}(c) shows a histogram of the xy crosstalk $\Lambda^\mathrm{(xy)}$ with average and standard deviation values of $-39.4\pm3.7\,$dB for the processor without tunnels, and $-37.4\pm3.9\,$dB for the processor with tunnels. These results show that the addition of the tunnel structures to shield the xy-lines do not decrease the average xy crosstalk for these packaged processors.

Figure~\ref{fig3}(d) shows a histogram of the dc-flux crosstalk $\beta^{\mathrm{(dc)}}$ measured both with and without the tunnel structures. Statistics from both data sets show average values of approximately 0.05\,\%.
Data from the processor without the tunnels has outliers in-between 0.3\,\% and 0.6\,\% (indicated as Group 2), which originate from pairs whose z-lines are not only nearest neighbour but also arranged in a specific way (see section \ref{main_section_distancescaling}). 
These outliers are not present in the data from the processor with the tunnel structures, indicating a suppression of dc-flux crosstalk; here, the maximum measured value is 0.13\,\%. We also note that the measured dc-flux offsets from the processor with tunnels ($\Phi_\mathrm{offset, tunnel}/\Phi_0=-0.035\pm0.008$) are more narrowly distributed than those from the processor without tunnels ($\Phi_\mathrm{offset, no-tunnel}/\Phi_0=-0.048\pm0.044$, excluding an outlier at $0.325\,\Phi_0$).

As the ac-flux crosstalk $\beta^{\mathrm{(ac)}}$ measurement is more involved, we only characterised it for Group-2 pairs of z-lines and couplers, i.e. those that exhibited elevated dc-flux crosstalk in the absence of tunnels. 
We chose to characterise it at a signal frequency of $200$\,MHz, close to the typical modulation frequency of our parametric couplers. Similarly to dc-crosstalk, the largest ac-crosstalk drops from 0.85\,\% to 0.24\,\% for tunnel-covered z-lines.

In the next section, we examine the distance dependence of the $\Lambda^\mathrm{(xy)}$, $\beta^{\mathrm{(dc)}}$, and $\beta^{\mathrm{(ac)}}$ data.

\section{Crosstalk vs.\@ distance}\label{main_section_distancescaling}

\begin{figure}
\centering  \includegraphics{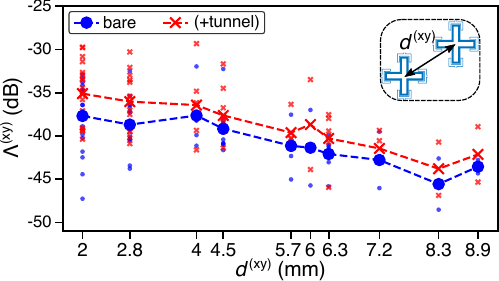}
  \caption{Distance dependence of the xy crosstalk, $\Lambda^\mathrm{(xy)}$. The distance $d^\mathrm{(xy)}$ is defined as the separation between the victim qubit and the target qubit of the source xy-line as illustrated in the inset, and further described in Section \ref{main_section_distancescaling}. The scatter plot indicates the distribution of $\Lambda^\mathrm{(xy)}$ data for each $d^\mathrm{(xy)}$, and the larger symbols denote the average.
  } \label{fig4}
\end{figure}

\begin{figure}
\centering
\includegraphics{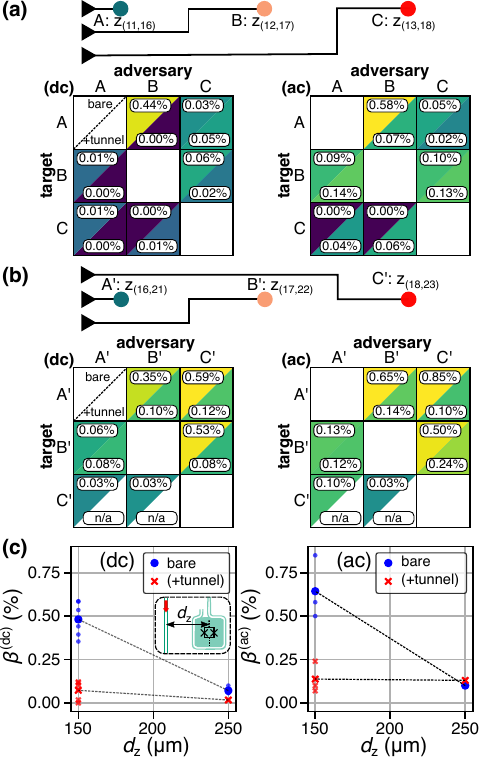}
  \caption{Layout-dependent flux-crosstalk behaviour. (a,b) Detailed comparisons of flux crosstalk for two different configurations of nearest-neighbour z-lines. The darker the background colour, the smaller the crosstalk. 
  (c) Distance dependence of flux dc and ac crosstalk for pairs of source--victim arranged as shown in the inset, where the red arrow indicates the source current. The dashed lines connecting the data points are meant to guide the reader's eyes.}\label{fig5}
\end{figure}

During the operation of a quantum processor, gate operations on different qubits and couplers will be performed simultaneously via multiple signal lines. 
It is therefore important to design the QPU such that the crosstalk decreases for larger separation. Such a property allows reusing the qubit parameters from one unit cell in another as part of our qubit-frequency allocation scheme; see fig.\,\ref{fig2}(f). It also ensures that the total error due to signal crosstalk remains sufficiently small. 
In this section, we begin to address this by demonstrating smaller average crosstalk for larger separations.

We sorted the $\Lambda^\mathrm{(xy)}$ data in fig.\,\ref{fig3}(c) according to the distance $d^\mathrm{(xy)}$ between the victim qubit and the target qubit of the source xy-line; the results are shown in fig.\,\ref{fig4}. It should be pointed out that there are fewer data points for larger $d^\mathrm{(xy)}$ in fig.\,\ref{fig4}, i.e this is natural for the planar connectivity such as the one implemented in our processors. The dashed lines connecting average $\Lambda^\mathrm{(xy)}$ values ($\bar{\Lambda}^\mathrm{(xy)}$, indicated by larger symbols of circles and crosses) hint at a trend of decreasing $\bar{\Lambda}^\mathrm{(xy)}$ with $d^\mathrm{(xy)}$. Fitting the average values with an empirical linear model (see Appendix \ref{app_empiricalfit}) yields a slope of approximately $-1\,$dB/mm for these processors.

For the flux crosstalk data associated with neighbouring z-lines, we measured a total of 4 different configurations within the two lowest routing corridors shown in fig.\,\ref{fig2}(b). Here, we focus specifically on two layouts of neighbouring z-lines as shown in fig.\,\ref{fig5}(a,b), but the trend is similar in the other two. 
The first layout consists of the set \{A, B, C\} = \{z$_\mathrm{(11,16)}$, z$_\mathrm{(12,17)}$, z$_\mathrm{(13,18)}$\}, and the second consists of \{A', B', C'\} = \{z$_\mathrm{(16,21)}$, z$_\mathrm{(17,22)}$, z$_\mathrm{(18,23)}$\}. 
The two layouts are very similar: most of the lines are nearest neighbours, beginning from the wirebond launchpads and leading all the way to the target position (see chip layout in fig.\,\ref{fig2}). 
The only major difference is in the third z-line: C' is routed on the \textit{upper} side of A' and B', while C is routed on the \textit{lower} side of A and B. 
The data with C' as victim from the processor with tunnels is not available due to a defective SQUID in coupler cp$_\mathrm{(18,23)}$.

On the first layout without the tunnel structure, fig.\,\ref{fig5}(a), the highest crosstalk $\beta$ (dc: 0.44\,\%, ac: 0.58\,\%) appears with A as the victim and B as the source line. 
However, a much lower crosstalk (dc: 0.06\,\%, ac: 0.10\,\%) appears between victim B and source C despite being nearest-neighbour lines in the same way as victim A and source B.
Almost similar results are observed on the second layout in fig.\,\ref{fig5}(b), with the exception that substantial crosstalk is also observed between victim A' and source C' (also between 
victim B' and source C'). 

Another noteworthy feature is the asymmetry of crosstalk for nearest-neighbour lines. 
The effect is most pronounced for the processor without tunnels. 
Whereas there is a substantial crosstalk from source B to victim A (dc: 0.44\,\%, ac: 0.58\,\%), a much lower crosstalk is measured when the roles of victim and source are interchanged (dc: 0.01\,\%, ac: 0.09\,\%). 
Similar asymmetry is observed in the other layout.

These observations, combined with the symmetric positioning of the wirebond launchpads of the z-lines, indicate that the major source of flux crosstalk in the processor without tunnels is coming from the proximity of the end (current loop) of a victim z-line and the nearest section of a source z-line. 
Under this model, we plotted flux crosstalk for pairs of victim--source arranged as shown in the inset of fig.\,\ref{fig5}(c). 
Without tunnels, the flux crosstalk becomes smaller for larger spatial separation $d_\mathrm{z}$. 
By adding tunnels, the previously high flux crosstalk at small separation ($d_\mathrm{z}=150\,\upmu$m) decreases by close to 10 times, down to the level of 0.1--0.2\,\%. 
Flux crosstalk at larger separation ($d_\mathrm{z}=250\,\upmu$m) is relatively unaffected by the presence of tunnels.

\section{Discussion}
\subsection{Qubit-drive crosstalk}
A distinct feature in this processor is the presence of densely routed signal lines in the proximity of qubits within the flip-chip environment; see fig.\,\ref{fig2}(b). 
However, for xy crosstalk, direct capacitive interaction between qubits and these signal lines cannot be the dominant mechanism, since we do not observe any decrease in average crosstalk after covering the xy-lines with shielding tunnel structures; see figs.\,\ref{fig3}(c) and \ref{fig4}. This conclusion is consistent with several electrostatic simulations designed to study the expected impact of the tunnel structure and xy-crosstalk level due to the direct xy-qubit capacitive interaction. First, we simulate a simplified model containing two victim qubits (denoted as q$_1$ and q$_2$) and two source xy-lines (denoted as xy$_3$ and xy$_4$) under two situations: with and without a tunnel structure on xy$_4$ (see Appendix \ref{app_tunneleffect}). The result shows that while the coupling capacitances between xy$_3$ and both qubits remain unaffected, the coupling capacitances between xy$_4$ and both qubits are lower when the tunnel structure is present on top of xy$_4$. Second, we simulate the expected xy-crosstalk level that is solely caused by the direct capacitive interaction between the xy-lines in the processor (excluding wirebonds, PCB signal traces, and tunnels) and the qubits (see Appendix \ref{app_xyqubitdirectcrosstalk}). The result puts the expected xy-crosstalk level within the range of $-49\,$dB to $-150\,$dB, much smaller than the measured xy-crosstalk level in our work.

The average of $\Lambda^\mathrm{(xy)}$ is $-36$\,dB for the nearest-neighbour pairs ($d^\mathrm{(xy)}=2.0\,$mm) and $-37$\,dB for the next-nearest-neighbour pairs ($d^\mathrm{(xy)}=2.8\,$mm). 
The highest values are $-30$\,dB for both nearest and next-nearest neighbours.
This performance compares favourably with results from a 17-qubit processor (in a single-chip geometry) employed in Ref.\,\cite{krinner2022} for the demonstration of distance-three quantum error correction. 
There, the highest crosstalk among the reported values for next-nearest-neighbour qubit ($d^\mathrm{(xy)}=2.4\,$mm) was $-20$\,dB (excluding one outlier at $-4$\,dB). 
It is not straightforward to identify the factors that led to the lower xy-crosstalk level demonstrated in our work despite having a denser signal-line layout. 
However, there are good reasons to believe that the chip geometry is important. 
The electric field of a coplanar waveguide is more confined in a flip-chip geometry than in a planar chip, due to the presence of a ground plane on the opposing chip (the typical size of a transmission line is $\sim\!10\,\upmu$m, and our chip separation is $8\,\upmu$m) \cite{kosen2022}.
In a single planar chip, additionally, the signal lines can interact parasitically with qubits via the field in the substrate, whereas in our flip-chip module, signal lines and qubits are separated onto different chips.

The data for $\Lambda^\mathrm{(xy)}$ are sorted according to qubit-qubit distance $d^\mathrm{(xy)}$ in fig.\,\ref{fig4} purely for a scaling argument. 
We believe that for a signal delivery architecture to be scalable, not only should $\Lambda^\mathrm{(xy)}$ decrease with $d^\mathrm{(xy)}$, but the behaviour should not be too dependent on the architecture itself. 
Such a behaviour would not only unlock favourable scaling behaviour, it would also simplify the hardware design process by reducing the number of factors that need to be accounted for. 
When combined with a judicious choice of qubit-frequency allocation and a substantial decrease in $\Lambda^\mathrm{(xy)}$ for increasing $d^\mathrm{(xy)}$, it can lead to very low total crosstalk error. Having said that, the relatively large spread of $\Lambda^\mathrm{(xy)}$ for every qubit-qubit distance $d^\mathrm{(xy)}$ in fig.\,\ref{fig4} provides a clue that the major parasitic interaction in our processors does not have a strong dependence on $d^\mathrm{(xy)}$. 
The dominant source of crosstalk must therefore be sought elsewhere. 
Nevertheless, there is a hint of lower crosstalk for larger separation when comparing the average value of $\Lambda^\mathrm{(xy)}$ versus $d^\mathrm{(xy)}$, suggesting a favourable scaling behaviour, at least for the size of processors examined in this work.

An alternative to flip-chip geometry called the ``coaxmon'' reported nearest-neighbour ($d^\mathrm{(xy)}=2\,$mm) crosstalk $\Lambda^\mathrm{(xy)}\sim\!-56\,$dB, roughly 20\,dB better than the result reported in this work \cite{spring2022}. 
This was obtained using a smaller $2\times2$ qubit chip ($5\times 5\,$mm$^2$) in an enclosure that is inductively shunted at the center to repel the cavity modes to higher frequencies, an additional feature that is yet to be present in our system. It will be interesting to compare crosstalk performance between the two geometries for increasing processor size.

Going forward, a combination of experimental investigations on more dedicated devices and numerical validations will be useful to further understand various xy-crosstalk mechanisms and configurations in which they might prevail. Electromagnetic modelling of the whole system (chips, wirebonds, packaging), such as the one presented here, is computationally challenging due to the wide range of feature sizes (several micrometers for signal lines to several centimeters for the packaging) that need to be accounted for.

In this work, we have instead taken an alternative approach of simulating simplified versions of the model to understand the relative impact of direct capacitive interaction and the signal-lines proximity to the xy-crosstalk level. As discussed before and in more detail in Appendix \ref{app_xyqubitdirectcrosstalk}, the expected xy-crosstalk level caused by the direct xy-qubit capacitive interaction between the qubits and the signal lines (excluding the wirebonds, PCB signal traces, and tunnels) is between $-49\,$dB and $-150\,$dB with an average of $-94\,$dB, which is much smaller than the range measured in this work. In addition, we simulate the expected microwave crosstalk between neighbouring signal lines in a simplified model consisting of five signal lines beginning at the PCB level all the way into the flip-chip environments (Appendix \ref{app_linecrosstalk}). The results show nearest-neighbour and next-nearest neighbour crosstalk at the level of $-55\,$dB and $-69\,$dB at the PCB level, which are then substantially increased to $-40\,$dB and $-49\,$dB after including the wirebonds and signal launchpads. Further extension of the signal lines into the flip-chip environment does not substantially change the crosstalk level, a trend that is attributed to the much smaller geometry of the transmission line compared to the line-to-line separation. Crucially, the closest pair of xy-lines in our processor is at least in a \textit{next-nearest neighbour} configuration, as there is at least one z-line needed to control the tunable coupler connecting two neighbouring qubits. The simulation puts a bound of around $-47\,$dB to the next-nearest neighbour signal lines, which is still one order of magnitude \blue{smaller} than the average xy-crosstalk measured for qubits controlled by the next-nearest neighbour pairs of xy-lines ($-36\,$dB).

The xy-crosstalk performance of our quantum processors compares competitively with those demonstrated in trapped ions \cite{warring2013, piltz2014, audecraik2017, wang2020, pogorelov2021, fang2022, binai-motlagh2023a}, neutral atoms \cite{xia2015, levine2019, wang2016}, and spin qubits \cite{undseth2023, lawrie2023}.
In trapped-ion systems, the lowest nearest-neighbour crosstalk (without active cancellation) was reported to be 0.2\,\% for a linear chain of four $^{171}$Yb$^+$ ions, which is equivalent to an individual excitation-error probability of $10^{-5}$~\cite{wang2020}. 
In our case, the qubit-frequency allocation strategy results in nearest-neighbour qubits that are naturally far detuned. For $\Omega_\mathrm{R}/2\pi=25\,$MHz, and with the smallest nearest-neighbour detuning being $420\,$MHz, the individual excitation-error probability is already at the level of $10^{-6}$; larger detuning will further reduce the signal-crosstalk-induced error probability.
Neutral-atoms systems have reported an average spin-flip crosstalk probability of $2\times10^{-3}$ on a two-dimensional array of 49 Cs atoms \cite{xia2015}. 
Spin-qubit systems, such as the four-atom Ge processor in Ref.\,\cite{lawrie2023}, show an average individual single-qubit gate error between $10^{-2}$ and $10^{-4}$. 

\subsection{Flux crosstalk}
The flux-crosstalk data [fig.\,\ref{fig5}(a,b)] clearly illustrate specific configurations of z-lines that cause some of the largest flux-crosstalk values measured in our processors (between 0.4 and 1\,\%). We demonstrated two strategies to achieve lower flux crosstalk. The first strategy uses a tunnel structure to suppress crosstalk for the same pairs of z-lines below $\sim\!0.2\,\%$. This conclusion is only made possible due to the use of identical signal-line footprints in the two devices. The second strategy is by ensuring a sufficiently large separation $d_\mathrm{z}$ between the head of the victim z-line and the closest point of a source z-line, in our case $d_\mathrm{z}=250\,\upmu$m. Further investigation is needed to pinpoint precisely the cause of crosstalk reduction due to the tunnel structure: whether the tunnel prevents current leakage to the victim z-line and/or if it suppresses the direct interaction between the source z-line and the victim SQUID.

The average dc-flux crosstalk $\beta^\mathrm{(dc)}$ over all measurable pairs is $\sim\!0.05\,\%$. At least another tenfold reduction is possible through active flux compensation, potentially pushing the dc-flux crosstalk level below 0.01\,\%. A similar reduction has also been demonstrated in a 16-qubit processor \cite{barrett2023}. Note that the passive stability of our system is much better: a variation of $32\,\upmu\Phi_0$ is measured for one of the z-lines over a period of 3 hours, corresponding to a lowest measurable $\beta^\mathrm{(dc)} = 0.003\,\%$. See Appendix \ref{app_meastechnique_dc} for more information.

The relative standard deviation of dc-flux offset in the processor with the tunnels ($\pm0.008\Phi_0, 23\%$) is smaller than the value in the processor without the tunnels ($\pm0.044\Phi_0, 92\%$). In our system, we believe that the dc-flux offsets are determined by the magnetic field present in the system when the aluminium film becomes superconducting. The magnetic field can be due to the remnant static field generated by sources external to the processor as well as any circulating current in it. The two processors are measured in the same cryogenic environment, except for the different PCBs, oxygen-free high thermal conductivity copper base plate and cover lid, and screws. The remnant static field should, in principle, remain the same at the base temperature, and this is likely reflected in the non-zero average values. In addition, the input current into each z-line is always set to zero during the cool-down. Therefore, it is not straightforward to determine the role of the tunnel structure in narrowing down the dc-flux offset distribution solely from the data presented in this work.

There is a varying degree of tolerable $\beta^\mathrm{(dc)}$ performance. An active flux-calibration technique applied to devices for quantum-annealing applications achieves maximum calibration error of $0.17\,\%$ \cite{dai2021}, which is still higher than the passive performance already obtained by our architecture. In another work employing a superconducting quantum simulator to study many-body dynamics, $\beta^\mathrm{(dc)}$ was reduced from 6\,\% down to $0.01\,\%$ through active compensation \cite{zhang2023}. 

Overall, this work demonstrates that careful routing and shielding of z-lines can already enable low crosstalk without active flux compensation. The crosstalk performance can be compared with various leading devices (flip-chip, single-chip) where the maximum flux crosstalk (no compensation) varies between 0.8\,\% and 40\,\% \cite{neill2018,abrams2019,dai2021,krinner2022,braumuller2022,karamlou2023,zhang2023}. Authors in Refs.\,\cite{braumuller2022,karamlou2023} noted an improvement in maximum flux crosstalk down to the level of 1.6\,\% when moving from single-chip to flip-chip geometry on nominally similar chip sizes, attributed to the use of a smaller SQUID area. Meanwhile, Ref.\,\cite{krinner2022} already registered a maximum flux crosstalk of $0.8\,\%$ on a single-chip geometry, lower than those achieved in Ref.\,\cite{karamlou2023}. The use of airbridge crossovers meant to stitch the ground plane of a z-line did not immediately result in full crosstalk reduction as demonstrated by Ref.\,\cite{neill2018}, where the authors measured flux crosstalk between 0.1\,\% and 4\,\%. Finally, we note that Ref.\,\cite{zotero-4498} obtained $\beta^\mathrm{(dc)}=0.8\,\%$, despite using a z-line with a dedicated return-current line. This illustrates the complex story of flux-crosstalk improvement efforts, and the signal-line architecture presented in this work is a proof that lower crosstalk is still possible to achieve via careful design. Further investigations are required to understand the source of flux crosstalk at the 0.1\,\% level. 

\subsection{Impact on gate fidelities}
It is instructive to understand the detrimental effect of the xy-crosstalk performance on the single-qubit gate fidelity. In previous work, we calculated the fidelity using the relative coupling strength between the relevant processes and detuning between the associated transitions \cite{osman2023}. The calculation models the qubits as two-level systems and assumes that the crosstalk activates undesired processes one by one. The measured $\Lambda^\mathrm{(xy)}$ varies between $-56$\,dB and $-27$\,dB, which corresponds to relative coupling strengths between $0.15$\,\% and $4$\,\%. Assuming a $20$\,ns single-qubit gate, the worst $\Lambda^\mathrm{(xy)}$ of $-27$\,dB would require detuning between the relevant qubit transitions larger than $28$ and $42$\,MHz (about $1$\,\% of typical qubit frequencies) to have average gate fidelities above $99.9$\,\% and $99.99$\,\% respectively. 

The limits are more stringent when we consider influence from all other qubits on the chip. We examine the total single-qubit gate fidelity $F_{1Q}$ for increasing processor size. 
We assume an empirical linear behaviour of $\bar{\Lambda}^\mathrm{(xy)}$ on $d^\mathrm{(xy)}$ and numerically simulate the impact on $F_{1Q}$ under certain assumptions (see Appendices \ref{app_empiricalfit}, \ref{app_1qerror}, \ref{app_numericalsimulations}). 
For the data in fig.\,\ref{fig4}, the slope $m_\mathrm{xy}$ and intercept $\Lambda_0$ values of the empirical linear model are approximately $-1$\,dB/mm and $-34$\,dB, respectively. 
The simulations suggest, for QPUs with up to 1000 qubits, that an improvement to $-2.0$\,dB/mm and $-50$\,dB would result in a total single-qubit gate error (due to xy crosstalk) well below the 0.1\,\% threshold recommended for quantum error correction using the surface-code~\cite{martinis2015}. This conclusion assumes linear dependence of $\bar{\Lambda}^\mathrm{(xy)}$ vs.\@ $d^\mathrm{(xy)}$ for processor sizes beyond the one measured in this work. We have no reasonable physical argument to assume that it will be the case; nevertheless, this number can serve as a guide for future hardware development and can hopefully spur further effort in verifying this relation for larger processors. Note that a sparser lattice architecture, such as the heavy square or the heavy hexagon, will benefit from less stringent criteria due to fewer neighbouring qubits and more relaxed frequency constraints~\cite{hertzberg2021}.

It is more challenging to analyze the influence of flux crosstalk on the two-qubit gate fidelity since the gate strength depends on both the dc bias and the ac amplitude in a nontrivial way ($\omega_\mathrm{cp}(\Phi)$ is non-linear). However, there is a relevant situation for which the analysis becomes simpler, at least qualitatively: a parametric gate is applied to a coupler while the other ones idle. Two related conditions arise in that case: (1) the driving amplitude produced on the idling couplers will be very small for low crosstalk such as the one measured in this work, allowing for a perturbative expansion, and (2) the dc contribution stemming from the periodic modulation will be negligible, as it is a higher-order effect in the small-amplitude perturbative expansion. Furthermore, as the dc bias usually stays untouched during the parametric-gate operation, the effect of dc crosstalk on the idling couplers should not change during the gate operations.

An expression for the first-order contribution to the gate strength of the iSWAP gate is known for this small-amplitude expansion \cite{roth2017}, although the result is limited to the regime where the detunings between the coupler and the qubits are much larger than their coupling strengths. The expression clearly shows that the gate strength is less sensitive to changes in both ac amplitude and dc bias near the zero-flux bias region. This region is also where the couplers are typically more far-detuned from the qubits, thus avoiding the physical consequences of being near avoided level crossings, i.e. Rabi oscillations or extra phase acquisition. Therefore, the general prescription to avoid big impact of crosstalk during the application of parametric gates would be to idle the other couplers to be as far detuned as possible from the neighboring qubits. 

\section{Conclusion and outlook}
We demonstrated superconducting quantum processors utilising an on-chip signal-delivery architecture with competitive crosstalk performance --- average on-resonant xy crosstalk $\sim\!40\,$dB, average dc-flux crosstalk $\sim\!0.05\,\%$ --- surpassing a majority of those demonstrated in other physical platforms. The systematic comparison enabled by adding a tunnel structure on an identical signal-line footprint combined with electrostatic simulations show that the direct capacitive interaction between densely routed signal lines and the qubits are not the main contributor to the xy-crosstalk level in these processors. Electromagnetic simulation of the microwave crosstalk between neighbouring signal lines indicates a crosstalk level that is still one order of magnitude smaller than our measured xy-crosstalk level. Our work demonstrates wiring layouts that can lead to flux crosstalk at the $\sim1$\,\% level and strategies to reduce it down to $\sim0.1\,\%$. 

Further investigations into the source of crosstalk would be required before proposing reliable mitigation strategies; the lists below are non-exhaustive. For xy crosstalk, there are at least three possible avenues to investigate. The first is to numerically investigate the additional xy crosstalk caused by the direct capacitive interaction between a victim qubit and the PCB signal traces including the wirebonds. While they are further away from the qubits, they are also much larger in size compared to the on-chip signal lines. The second is to investigate the role of packaging cavity modes in influencing the xy-crosstalk level as suggested in Ref.\,\cite{wenner2011a}, in particular the indirect xy-qubit interaction mediated by the cavity modes. The third is to investigate the indirect xy-qubit interaction mediated by other elements in the processor such as the qubits, the couplers, the readout resonators, and possible parasitic modes. For flux crosstalk, it will be instructive to investigate if it is limited by the leakage of the return current into the victim line (around the launchpads or the area around the SQUID), or if the return current is contributing non-negligible magnetic flux at the location of the victim SQUID.

More advanced signal-delivery architectures will benefit from the use of on-chip signal multiplexers, bumps, and especially redistribution layers enabled by through-silicon vias (TSV) in a multi-tiered stack \cite{vahidpour2017, yost2020, mallek2021a, grigoras2022, hazard2023}. Bumps and TSVs are typically a few tens of micrometer in size, which is relatively large compared to typical transmission lines. It remains to be seen the extent to which they influence the overall signal-crosstalk level.

The low crosstalk demonstrated in this work and other cited works highlights the strength and flexibility afforded by superconducting circuit architectures. There is still no clear lower limit for the achievable passive crosstalk performance via further system design. We hope to inspire further efforts and discussion in the quantum hardware community, both within and outside superconducting platforms, to investigate the source of crosstalk as well as its behaviour as systems scale up in size.

\section*{Acknowledgements}
This work was funded by the Knut and Alice Wallenberg (KAW) Foundation through the Wallenberg Centre for Quantum Technology (WACQT) and the European Union Flagship on Quantum Technology HORIZON-CL4-2022-QUANTUM-01-SGA project 101113946 OpenSuperQPlus100. We are grateful to Eleftherios Moschandreou for the technical support.
We acknowledge the use of support and resources from Myfab Chalmers, the National Academic Infrastructure for Supercomputing in Sweden (NAISS), and the Swedish National Infrastructure for Computing (SNIC) at Link\"{o}ping University (partially funded by the Swedish Research Council through grant agreements no.\,2022-06725 and no.\,2018-05973). 

\section*{Author Contribution}
SK planned and performed the experiments, analysed the data, and produced the numerical simulations. 
SK and HXL designed the QPU with inputs from  MR, RR, JFP, AO, AFR, DS, AFK, GT, and J.Byl.
MR, SK, AN, and HXL fabricated the control and quantum chips with inputs from J.Biz, AO, and AFR.
MC and LG formed the In bumps and performed the flip-chip bonding.
RR, GT, and SK developed the QPU packaging.
CK assisted with the writing of measurement instrument drivers.
SK, LC, CK, HXL, and GT set up and maintained the measurement and cryogenic facilities.
SK and J.Byl wrote the manuscript with input from all authors. GT, JG, and J.Byl supervised the project.

\section*{Competing Interests}
RR, GT, and SK are co-founders and shareholders of Scalinq AB. Other authors declare no competing interests.

\section*{Data Availability}
The data that support this work are available from the corresponding author upon reasonable request.

\appendix
\section{Fabrication and packaging\label{app_fabricationpackaging}}
In this section, we outline the fabrication and packaging steps of the processors used in this work. The overall procedure, with the exception of the tunnel structure fabrication, closely follows Refs.\,\cite{kosen2022, chayanun2024, burnett2019}.

The first fabrication steps are done on our two-inch wafers at Chalmers University of Technology. We use two high-resistivity intrinsic silicon wafers: one containing four designs of C-chip, and another one has four designs of Q-chip. The fabrication process of the ground and wiring layer follows the `Chalmers standard fabrication process' outlined in Ref.\,\cite{chayanun2024}. Each wafer goes through the \textit{SC-1} process, then a deionised(DI)-water rinse, followed by a one-minute dip in a $2\%$ aqueous solution of hydrofluoric acid, before another round of DI-water rinse. The wafer is blow-dried using nitrogen gas and is immediately loaded into the load-lock chamber of an electron-beam evaporator to minimise the re-oxidation of the silicon surface. Inside the vacuum chamber, the wafer is heated to $300^\circ$ Celcius for ten minutes, and is let to cool down over the course of $\sim\!20$ hours. Then a $150\,$nm thick aluminium film is evaporated on the wafer followed by static oxidation of the exposed aluminium surface for ten minutes before we unload the wafer from the evaporator.

Next, a $50\,$nm thick niobium nitride film is sputtered on electron-beam patterned resists (after an in-situ ion milling step to remove the oxide layer on top of the aluminium film), followed by a liftoff process to create the underbump metallisation layer on each wafer. Then the ground and wiring layer is formed on the aluminium film using optically patterned resists followed by a wet etching step.

On the wafer containing the Q-chips, we form cross-type Josephson junctions using a two-step process similar to Ref.\,\cite{burnett2019}. The first step is to pattern the junction electrodes on a resist stack using the electron-beam lithography, followed by a two-angle evaporation technique (aluminium junction) in an electron-beam evaporator, then a lift-off process. The second step is to form the patch layer connecting each junction with the rest of the qubit circuitry. The patch layer is electron-beam patterned on a resist stack, followed by ion milling the exposed wiring layer and aluminium film deposition, after which we perform a lift-off process.

In parallel, we form the tunnel structures following Ref.\,\cite{chen2014}, on the wafer containing the C-chips. First we optically define a pattern of the tunnel `foot' across the wafer on a resist layer, which is then reflowed to form the arch of the tunnel. Then we perform ion milling on the exposed ground plane before evaporating another aluminium film for the tunnel structure. At this point, the tunnel foot region is the only place where the aluminium tunnel structure is in contact with the ground plane of the chip. Next, we add a second resist on top of the wafer and optically expose the area that is not part of the tunnel structure. The wafer then goes through a wet etching process before we remove the rest of the resists from the wafer.

We package and send both wafers in an anti-static bag to VTT. There, $8\,\upmu$m-tall indium pillars are formed on both wafers on an optically patterned resist layer with a side profile optimised for lift-off. The wafers are then diced into chips before being flip-chip bonded at room temperature. The chip separation is characterised by measuring the distance between the two chips at the the various edges of the flip-chip module using a scanning electron microscope. The flip-chip modules are then sent back to Chalmers.

At Chalmers, we glue the flip-chip module / processor on the four corners using GE varnish, and let it dry overnight. Afterwards, aluminium wirebonds are applied using an automatic wirebonding machine. The packaged module is finally installed at the mixing chamber stage of the dilution fridge.

\section{Experimental setup}
\begin{figure}[t]
\centering
\includegraphics{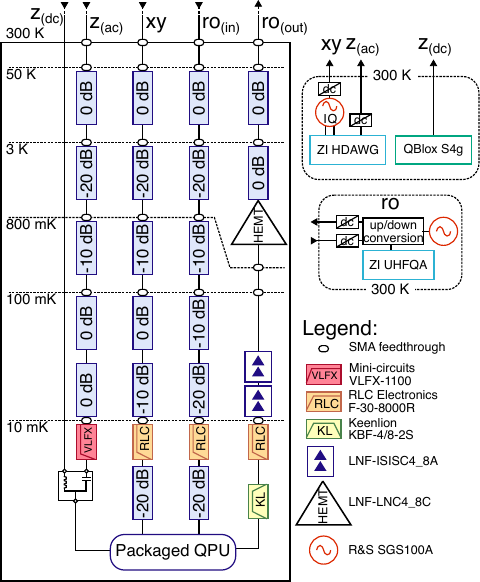}
  \caption{Measurement apparatus at cryogenic and room temperature.} \label{app_fig_apparatus}
\end{figure}

Figure \ref{app_fig_apparatus} shows the measurement apparatus employed in this work at cryogenic and room temperature. 

\section{Measurement techniques} \label{app_meastechnique}
\subsection{xy crosstalk}
\begin{figure}[t]
\centering
\includegraphics{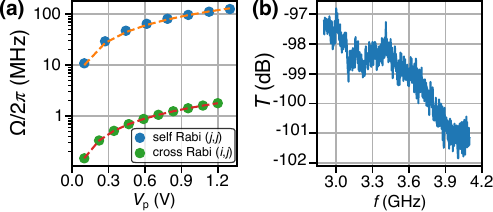}
  \caption{xy-crosstalk measurement. (a) Measured Rabi frequencies as a function of the drive amplitude for the self-drive ($i=j$) and cross-drive ($i\ne j$). (b) An example of combined output and transmission characteristics of an rf source with the cables and components all the way down to the PCB level (measured at room temperature).} \label{app_fig_xycrosstalk}
\end{figure}

The xy crosstalk $\Lambda_{i,j}$ [eq.\,(\ref{eq:xy_crosstalk_definition}) in the main text] is the ratio of the on-resonant Rabi frequencies squared, i.e $(\Omega_{i,j}/\Omega_{j,j})^2$, expressed in decibel for a fixed drive amplitude $V_j$; see fig.\,\ref{fig3}(a).  We obtained a series of Rabi frequencies for various drive amplitudes [fig.\,\ref{app_fig_xycrosstalk}(a)] and fitted the data with a linear relation $\Omega_{i,j}=k_{i,j}V_j$ to obtain the slope $k_{i,j}$. We obtained $k_{j,j}$ in a separate measurement. $\Lambda_{i,j}$ is then equal to the ratio $(k_{i,j}/k_{j,j})^2$ expressed in decibel.

We operated the radio-frequency (rf) source (Rohde \& Schwarz SGS100A) in IQ modulation mode. We fixed its output power to $14$\,dBm and varied the amplitudes $V_j$ of the signals sent to its IQ input ports. For each $V_j$, we obtained a Rabi oscillation and fitted it with an exponentially decaying sinusoidal function. Figure \ref{app_fig_xycrosstalk}(a) shows $\Omega_{i,j}$ and $\Omega_{j,j}$ vs.\@ $V_j$ for $i=15,\,j=14$. 

In the measurement of $\Lambda_{i,j}$, we took into account the frequency dependence of the transmission of the cables and components, measured using a vector network analyser at room temperature. 
We measured the output characteristic of the rf source for different frequencies with a spectrum analyser (Rohde \& Schwarz FSL18, frequency span 1\,MHz, intermediate frequency bandwidth 10\,KHz, 20 averages, 101 points). 
Figure \ref{app_fig_xycrosstalk}(b) shows an example of the transmission characteristic of the cables and components.

\subsection{dc-flux crosstalk}\label{app_meastechnique_dc}
\begin{figure}[t]
\centering
  \includegraphics{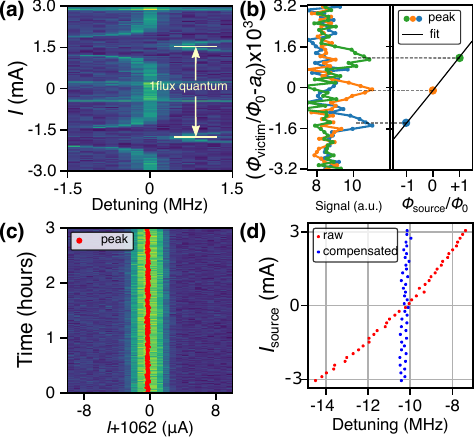}
  \caption{dc-flux crosstalk measurement. (a) Qubit- frequency spectroscopy vs.\@ current $I$ applied to a neighbouring tunable coupler. Detuning refers to the frequency difference between the probe tone and the qubit  at $I=0$. The avoided-crossing regions are used to probe for small shifts in the coupler frequency due to  qubit-coupler hybridization. The ``1 flux quantum'' label refers to the magnetic-flux periodicity of the coupler frequency. 
  (b) Shift in the flux $\Phi_\mathrm{victim}$ of the victim coupler vs.\@ flux $\Phi_\mathrm{source}$ of the source coupler. Each curve in the left panel represents a narrow vertical slice of the data in (a) around the vicinity of one of the avoided crossings ($a_0=0.365696$). For each $\Phi_\mathrm{source}$, we extracted $\Phi_\mathrm{victim}$ as shown in the right panel. The slope yields the dc-flux crosstalk coefficient (here $|\beta_{i,j}|=0.13\,\%$). (c) Repeated qubit-frequency spectroscopy for 3 hours to demonstrate the stability of the system. Here, we fixed the probe frequency and varied the current $I$ at the neighbouring coupler. (d) Demonstration of active dc-flux compensation by subtraction of the crosstalk.}
  \label{app_fig_dcfluxcrosstalk}
\end{figure}

The first step in the characterisation of flux crosstalk is to relate the z-line's direct current (dc) $I$ with the magnetic flux $\Phi^\mathrm{(dc)}$ applied to the SQUID of the coupler; see figs.\,\ref{fig2}(c) and \ref{fig3}(b). This is done via continuous-wave (cw) frequency spectroscopy of the neighbouring qubit (which is coupled to the victim coupler) as shown in fig.\,\ref{app_fig_dcfluxcrosstalk}(a). When the coupler comes close to resonance with the qubit, we observe an avoided crossing. By identifying a series of such crossings, we are able to deduce the current required to apply one flux quantum to the SQUID, as well as the dc offset current corresponding to the true zero dc-flux bias on the SQUID. In our case, one flux quantum ($\Phi_0=2.067\times 10^{-15}$\,Wb) corresponds to approximately 3\,mA. 

To measure dc-flux crosstalk $\beta^\mathrm{(dc)}_{i,j}$ [eq.\,(\ref{eq:z_crosstalk_definition}) in the main text], the victim coupler cp$_i$ is first biased close to one such avoided crossing. In the absence of a dedicated coupler-state readout resonator, this is a sensitive region to probe $\beta^\mathrm{(dc)}$ through qubit-state readout. 
A small shift in the frequency of cp$_i$, from $f_\mathrm{p}$ to $f'_\mathrm{p}$, can be inferred from the change in the hybridised coupler-qubit frequency $f_\mathrm{p}$, imparted by the source current $I_j$.
In an ideal condition with zero crosstalk, i.e. $\beta^\mathrm{(dc)}_{i,j}=0$, $f_\mathrm{p}$ is independent of the value of the current $I_j$ applied on source z$_j$. Due to non-negligible crosstalk, $f'_\mathrm{p}$ will slightly shift as a function of $I_j$, and the goal is to find the new current value $I'_i$ that restores $f'_\mathrm{p}$ to $f_\mathrm{p}$. In practice, we fixed the qubit probe frequency $f_\mathrm{p}$ (typically $\sim10\,$MHz away from the qubit bare frequency), and determined $I'_i$ for three different dc currents $I_j$ corresponding to source coupler (cp$_j$) flux bias values $-\Phi_0,\,0,$ and $\Phi_0$; see fig.\,\ref{app_fig_dcfluxcrosstalk}(b). We obtained $\beta^\mathrm{(dc)}_{i,j}$ by converting $I'_i$ to the corresponding flux applied on z$_i$. 

Figure \ref{app_fig_dcfluxcrosstalk}(c) demonstrates the passive stability of the dc-flux environment of a z-line for 3 hours. We repeatedly swept the input current to the z-line while maintaining the same probe frequency. From the positions of the extracted peak, we have obtained a standard deviation of the variation of the current to be $100$\,nA --- or, equivalently $32\,\upmu\Phi_0$ --- which is smaller than the minimum step in output current allowed by the source (approximately $380$\,nA). Furthermore this flux variation is much smaller than the typical dc-flux crosstalk observed in our system, which is around $500\,\upmu\Phi_0$ when $1\Phi_0$ is applied to the source z-line. 

We also demonstrate active dc-flux compensation on one of the victim z lines using the obtained crosstalk coefficient, as shown in fig.\,\ref{app_fig_dcfluxcrosstalk}(d). As the dc flux on the source line is varied between -$\Phi_0$ and $\Phi_0$ (the vertical axis is displayed in the equivalent unit of current), the range of variation in the resonance frequency (horizontal axis) drops more than tenfold (when compared to the uncompensated version, labeled as "raw" in the figure), from $7.12$\,MHz to $0.53$\,MHz. Further studies are required to understand this number and optimize the crosstalk-calibration procedure to push it below the limit allowed by the measurement setup

\subsection{ac-flux crosstalk}
For the case of ac-flux crosstalk, interference between ac-flux signals applied to the victim z-line and parasitic signals from the source z-line renders the technique used for dc-flux crosstalk measurement inapplicable. Instead, we employ a modified Ramsey pulse sequence following Ref.\,\cite{abrams2019}. We first give an overview of this technique before describing the calibration and measurement steps leading up to the crosstalk data.

\begin{figure}[t!]
\centering
\includegraphics{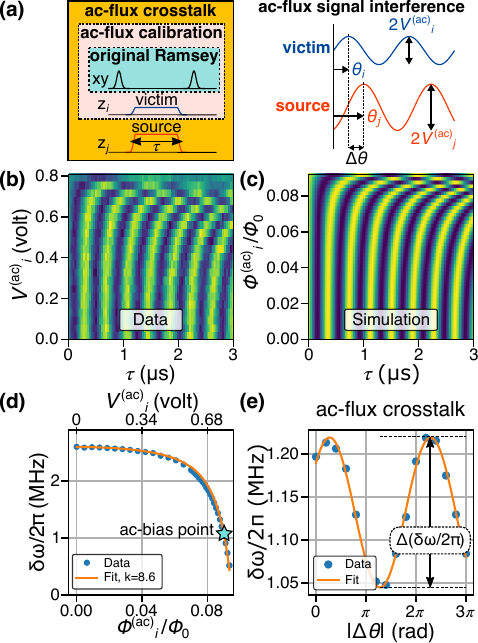}
  \caption{ac-flux crosstalk measurement. (a) Control pulse sequence. (b) Data obtained from running the ac-flux calibration sequence for a series of ac voltages $V_i^{\mathrm{(ac)}}$. (c) Simulated version of the data in (b) using the measured parameters of the device for a series of $\Phi_{i}^\mathrm{(ac)}$. (d) Fits of the extracted frequency $\delta\omega$ of the data in (b) to the simulation in (c), assuming a linear relation $V^\mathrm{(ac)}=k(\Phi^{\mathrm{(ac)}}/\Phi_0)$. The star symbol indicates the typical ac-flux bias point. (e) Measured $\delta\omega$ for $0<\Delta\theta<3\pi$ in the presence of source ac-flux. The oscillation amplitude (swing of $\delta\omega$) is proportional to the ac-flux crosstalk $\beta^\mathrm{(ac)}$ as discussed in the main text.} \label{app_fig_acfluxcrosstalk}
\end{figure}

First, consider a  qubit-coupler system (q$_i$, cp$_i$). The original Ramsey pulse sequence consists of two frequency-detuned $\pi/2$-pulses on the qubit, separated by an idle time $\Delta t$; see fig.\,\ref{app_fig_acfluxcrosstalk}(a). By measuring the q$_i$ population as a function of $\Delta t$, we obtain a Ramsey fringe whose oscillation frequency $\delta\omega=\omega^\mathrm{q}_i-\omega_\mathrm{drive}$ corresponds precisely to the detuning between the applied pulse and the actual frequency of q$_i$. In the modified Ramsey pulse sequence, an ac-flux pulse at angular frequency $\omega_{i}^{\mathrm{(ac)}}$ is also applied to the coupler, via z$_i$, during that idle duration, as shown in fig.\,\ref{app_fig_acfluxcrosstalk}(a). This ac-pulse modulates the qubit frequency and consequently also the detuning, which makes the observed $\delta\omega$ sensitive to the applied ac-flux pulse amplitude.

Now we modulate the victim coupler cp$_i$ through a source line z$_j$ to quantify the ac-flux crosstalk $\beta_{i,j}^\mathrm{(ac)}$. The victim coupler is biased at specific dc-flux and ac-flux bias points (angular frequency $\omega_{i}^{\mathrm{(ac)}}$, phase $\theta_i$) chosen to render it susceptible to detecting parasitic ac-flux signals. 
An ac-flux signal on a source line z$_j$, at frequency $\omega_{j}^{\mathrm{(ac)}}=\omega_{i}^{\mathrm{(ac)}}$, will cause a noticeable shift in $\delta\omega$ obtained via the modified Ramsey pulse sequence. 
We measure $\delta\omega$ for a range of phase offsets $\Delta\theta=|\theta_j-\theta_i|$ spanning $[0, 3\pi]$, as shown in fig.\,\ref{app_fig_acfluxcrosstalk}(e). The maximum swing of $\delta\omega$ (or $\Delta(\delta\omega)$) corresponds to an equivalent ac-flux shift in cp$_i$: $\Delta\Phi_{i}^{\mathrm{(ac)}}=2\beta_{i,j}^{\mathrm{(ac)}}\Phi_{j}^\mathrm{(ac)}$. 

In practice, we proceeded with the following steps: 
(1) relate $\Phi^{\mathrm{(ac)}}$ to the ac-voltage amplitude $V^{\mathrm{(ac)}}$ applied by the instrument, 
(2) make a choice of dc- and ac-flux bias points for the victim cp$_i$, 
(3) apply a large ac-flux signal on the source z$_j$, 
(4) run the sequence in fig.\,\ref{app_fig_acfluxcrosstalk}(a) and relate the measured swing $\Delta(\delta\omega)$ to $\Delta\Phi_{i}^\mathrm{(ac)}$ at the bias point, which in turn enables the determination of $\beta_{i,j}^\mathrm{(ac)}$.

To relate $V_{i}^\mathrm{(ac)}$ to $\Phi_{i}^\mathrm{(ac)}$ for any pairs of z$_i$ and cp$_i$, we ran the ac-flux calibration pulse sequence in fig.\,\ref{app_fig_acfluxcrosstalk}(a). Figure \ref{app_fig_acfluxcrosstalk}(b) shows the measured Ramsey fringes for a range of voltages $V_{i}^\mathrm{(ac)}$. In parallel, we ran a Qutip simulation \cite{johansson2012, johansson2013} of this pulse sequence using the measured parameters of qubit, qubit drive, and the coupler zero-flux bias frequencies. This simulation leads to equivalent Ramsey fringes for a range of $\Phi_{i}^\mathrm{(ac)}$; an example is shown in fig.\,\ref{app_fig_acfluxcrosstalk}(c). Finally, the extracted frequencies of the measured and the simulated Ramsey fringes are fitted to each other by assuming a linear relation, i.e. $V^\mathrm{(ac)}_i=k_i(\Phi^{\mathrm{(ac)}}_i/\Phi_0)$. In our setup, we typically obtain $k_i$ in the range from 6 to 10.

The choice of dc-flux bias points is made by following the typical bias points we use for two-qubit gate operation, which are around $0.3\Phi_0$ \cite{kosen2022}. The ac-flux bias point is generally chosen for maximum sensitivity to small changes in ac-flux; an example is shown in fig.\,\ref{app_fig_acfluxcrosstalk}(c). For the ac-flux frequency, we set $\omega^{\mathrm{(ac)}}/2\pi=200\,$MHz, which is a typical frequency for parametric modulation in this system.

Next we assume that the measured ac-flux crosstalk is small, which allows us to linearise the region around the bias point, i.e. $\Delta\Phi^{\mathrm{(ac)}}_i\approx\gamma^{\mathrm{(ac)}}_i\Delta (\delta\omega)$. The slope $\gamma^{\mathrm{(ac)}}_i$ is extracted using the nearest data points, and it allows us to infer $\Delta\Phi^{\mathrm{(ac)}}_i$ from the measured $\Delta( \delta\omega)$: $\beta_{i,j}^\mathrm{(ac)}=\gamma_{i}^\mathrm{(ac)}\Delta (\delta\omega)/2\Phi_{j}^\mathrm{(ac)}$.

\section{Target frequencies}\label{app_targetfrequencies} 
\begin{figure}[h]
\centering\includegraphics{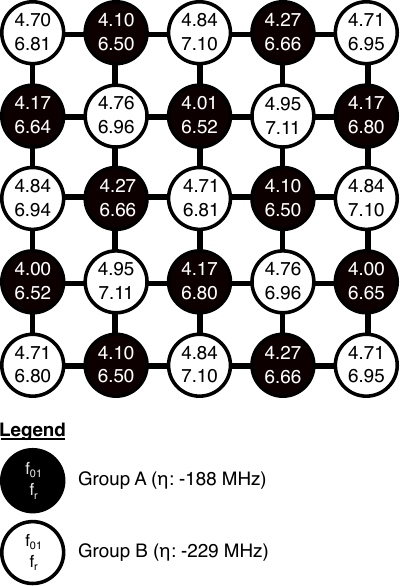}
  \caption{Target qubit frequencies ($f_{01}$, top) and readout resonator frequencies ($f_\mathrm{r}$, bottom). The alternating background colour illustrates the 2-frequency subgroup strategy with two distinct anharmonicity values ($\eta$).}\label{app_fig_targetfrequencies}
\end{figure}

The allocation of qubit frequencies in this processor follows the 2-frequency subgroup strategy as described in Ref.\,\cite{osman2023} and section \ref{main_quantumprocessorarchitecture} of the main text. The readout resonator frequencies are arranged to maintain qubit-resonator detunings that are above $2\,$GHz throughout the processor. All couplers are identical, and the target frequency at the zero-flux bias is $7.9\,$ GHz. Figure \ref{app_fig_targetfrequencies} shows the target qubit frequencies ($f_{01}$), anharmonicity ($\eta$) and the readout resonator frequencies ($f_\mathrm{r}$).

\section{Basic QPU parameters}\label{app_measuredparameters}
In table \ref{table_app_measfreq}, we list the range of measured or inferred qubit frequencies ($f_{01}$), anharmonicities ($\eta$), readout resonator frequencies ($f_\mathrm{r}$), coupler frequencies at zero flux bias ($f_\mathrm{c0}$), and single-qubit coherences ($T_1$, $T_2^*$). The qubits' and couplers' frequencies are approximately 1\,GHz lower than the design values due to an off-target Josephson-junction fabrication result.

\begin{table}[h]
\begin{tabular}{|>{\centering}p{2cm}|>{\centering}p{2cm}|>{\centering\arraybackslash}p{2cm}|} \hline
Parameters & QPU 1 & QPU 2 \\ \hline
 $f_{01}$ (GHz) & $[3.262, 4.077]$ & $[3.166, 3.982]$ \\ \hline
 $\eta$ (MHz) & $[-249, -191]$ & $[-245, -193]$ \\ \hline
 $f_\mathrm{r}$ (GHz) & $[6.317, 6.987]$ & $[6.335, 7.059]$ \\ \hline
 $f_\mathrm{c0}$ (GHz) & $[6.476, 7.049]$ & $[6.470, 7.329]$ \\ \hline
 $T_1$ ($\upmu$s) & $[76, 143]$ & $[30, 80]$ \\ \hline
 $T_2^*$ ($\upmu$s) & $[19, 83]$ & $[16,42]$ \\ \hline
\end{tabular}
\caption{Range of measured parameters of the two QPUs examined in this work. QPU 1 has the tunnels; QPU 2 does not have any tunnels.}\label{table_app_measfreq}
\end{table}

\section{Complete datasets}\label{app_completedatasets}
\begin{figure}[h]
\centering
  \includegraphics{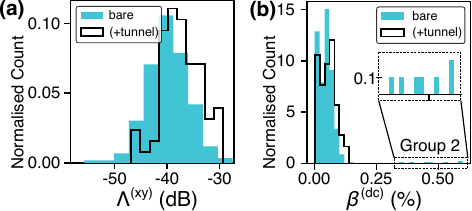}
  \caption{Complete datasets on the aggregate performance of xy crosstalk $\Lambda^\mathrm{(xy)}$ and dc-flux crosstalk $\beta^\mathrm{(dc)}$. (a) Histogram of the on-resonant xy crosstalk $\Lambda_{i,j}$ for a flip-chip module with bare transmission lines (labelled ``bare,'' 210 pairs, $\bar{\Lambda}_{\mathrm{bare}}=(39.8\pm4.0)$\,dB), and another flip-chip module with a majority of the transmission lines covered in tunnel structures (labelled ``+tunnel,'' 72 pairs,  $\bar{\Lambda}_{\mathrm{(+tunnel)}}=(37.4\pm3.9)$\,dB). (b) Histogram of the dc-flux crosstalk $\beta^\mathrm{(dc)}$ (bare: 462 pairs, tunnel: 274 pairs). Data indicated as Group 2 are associated with pairs whose z-lines are nearest neighbour.}\label{app_fig_completedataset}
\end{figure}

Figure \ref{app_fig_completedataset} shows the complete datasets of measured xy and dc-flux crosstalk from both processors. Due to the different numbers of pairs present in each dataset, we plot a normalised histogram indicated by the normalised count on the vertical axis.

\section{Expected effect of the tunnel}\label{app_tunneleffect}
\begin{figure}
\includegraphics{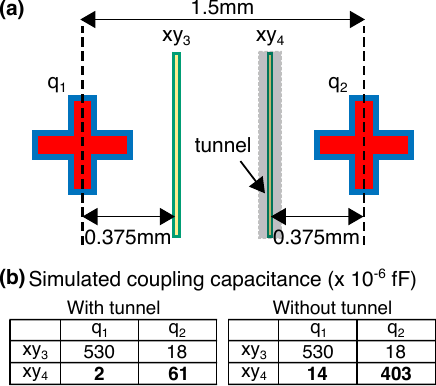}
\caption{Simulating the effect of the tunnel structure in reducing the direct xy-qubit coupling capacitance. (a) Simplified model of the simulation. Two qubit structures (q$_1$, q$_2$) on the Q-chip and two xy-lines (xy$_3$, xy$_4$) on the C-chip. The ground planes on both the C-chip and the Q-chip, and the coupler structure on the Q-chip, are not shown. (b) Extracted coupling capacitance values between two models: one with a tunnel structure on xy$_4$ and another one without a tunnel structure.}\label{app_fig_tunneleffect} 
\end{figure}

In this section, we are interested in the expected reduction of the coupling between a source xy-line and a victim qubit due to the tunnel structure. To do so, we simulated the coupling capacitance values between two xy-lines on the C-chip with two qubits on the Q-chip. Figure \ref{app_fig_tunneleffect}(a) shows a simplified version of the simulation model. The major differences with the actual processor layout are the qubit-qubit distance (which is $2\,$mm, in contrast to $1.5\,$mm in this model) and the closest distance between a source xy-line with a victim qubit (which is $500\,\upmu$m, in contrast to $375\,\upmu$m in this model). The tunnel structure is modelled as a metal enclosure with a sinusoidal roof profile ($3\,\upmu$m height, $50\,\upmu$m width) and no opening along both sides of the structure, which is different from the actual structure where there are periodically spaced openings along the tunnel to allow access to the cleaning solvent [see fig.\,\ref{fig2}(d) of the main text].
However, these differences should not, in principle, matter for the purpose of this simulation.

We use ANSYS Maxwell \cite{zotero-4428} to obtain the capacitance matrix. The table in fig.\,\ref{app_fig_tunneleffect}(b) shows the coupling capacitance values in two situations: with and without the tunnel structure on xy$_4$. First, the coupling capacitance is smaller for larger separation: compare $C(\mathrm{xy}_3, \mathrm{q}_1)$ and $C(\mathrm{xy}_3, \mathrm{q}_2)$ in both cases. As a reference, the coupling capacitance between a qubit and its own xy-line, e.g. $C(\mathrm{xy}_1, \mathrm{q}_1)$,  is at least 100 times larger than $C(\mathrm{xy}_3, \mathrm{q}_1)$. Second, by comparing the coupling capacitance values between xy$_4$ and either q$_1$ or q$_2$ before and after adding the tunnel structure, we can see that it does indeed weaken the direct coupling between the xy-line and the qubits. 
Note that $C(\mathrm{xy}_4, \mathrm{q}_2)< C(\mathrm{xy}_3, \mathrm{q}_1)$, despite both of them being identical in length and equally spaced from the respective qubits; this is because the CPW geometry of xy$_4$ is slightly narrower to retain $50\,\Omega$ characteristic impedance after adding the tunnel.

Therefore, we have shown that the tunnel structure can indeed reduce the direct capacitive coupling between an xy-line located on the C-chip and a qubit located on the Q-chip, at least for the simplified model discussed in this section. Given that the tunnel structure modifies the radiation profile of the xy-line, it is possible that other xy-crosstalk mechanisms may be impacted as well, and future work should investigate this in more detail.

\section{Crosstalk due to direct xy-qubit capacitive interaction}\label{app_xyqubitdirectcrosstalk}
In this section, we are interested in the expected crosstalk level due to the direct capacitive interaction between a victim qubit with any other source xy-line (excluding its own xy-line). The model includes both the full-sized C-chip and the Q-chip, without the wirebonds and the PCB packaging around it. To simplify the model, we retain the 25 transmon objects on the Q-chip and replace the coupler structures with a ground plane. On the C-chip, we only retain the 15 xy-lines controlling the three bottom rows of the qubits (q$_{11}$ to q$_{25}$). The rest of the xy-lines and z-lines are replaced with a ground plane. The model retains the full bump structures connecting the two chips. No tunnels are included in the model.

We employed ANSYS Maxwell \cite{zotero-4428} to simulate the capacitance matrix of such a problem. The relation between the capacitance values and the Rabi frequency ($\Omega_\mathrm{R}$) is obtained via the relation
\begin{align}
\int_0^{T_\pi}\Omega_\mathrm{R}(t) dt &= \pi, \\
2\sqrt{\kappa}bA_0T_\pi &= \pi,
\end{align}
where $\kappa$ is the photon loss rate via the xy-line, $b$ is the pulse-shape dependent constant obtained from the integration ($b=1$ for a square pulse, $b=2/\pi$ for a sinusoidal shape pulse), $A_0$ is the pulse amplitude, $T_\pi$ is the characteristic pulse duration to create a $\pi$-pulse. The expression for $\kappa$ is
\begin{align}
\kappa = Z_\mathrm{tml}\frac{C_\kappa^2\omega_\mathrm{q}^2}{C_\mathrm{q}},
\end{align}
where $Z_\mathrm{tml}$ is the xy-line characteristic impedance ($50\,\Omega$ in our case), $C_\kappa$ is the coupling capacitance between the xy-line and the qubit, $\omega_\mathrm{q}$ is the qubit angular frequency, and $C_\mathrm{q}$ is the qubit capacitance. This treatment follows closely the derivation described in Ref.\,\cite{blais2021} (specifically sections VIII.A, IV.F, IV.B, and appendix C in Ref.\,\cite{blais2021}). Reexpressing the xy-crosstalk definition in terms of the simulated capacitance values, we obtain
\begin{align}
\Lambda_\mathrm{dir}^\mathrm{(xy)}:=\Lambda_{i,j} & = 10\times\log_{10}{\left(\frac{\Omega_{i,j}}{\Omega_{j,j}}\right)^2}\\
& = 20\times\log_{10}{\left(\frac{\omega_iC_{i,j}}{\omega_jC_{j,j}}\sqrt{\frac{C_j}{C_i}}\right)},
\end{align}
where $C_{i,j}$ is the coupling capacitance between the victim qubit $i$ and the source xy-line $j$, and $\{C_i, C_j\}$ are the capacitances of qubit $i$ and $j$.

\begin{figure}
\centering\includegraphics{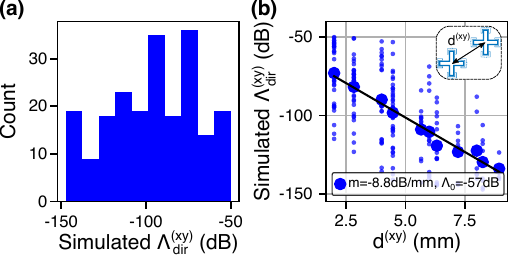}
  \caption{Simulation of xy crosstalk $\Lambda_\mathrm{dir}^\mathrm{(xy)}$ between victim qubits and other source xy-lines due to direct xy-qubit interaction. (a) Histogram of $\Lambda_\mathrm{dir}^\mathrm{(xy)}$ with an average and standard deviation of $(-96\pm26)\,$dB. (b) $\Lambda_\mathrm{dir}^\mathrm{(xy)}$ versus $d^\mathrm{(xy)}$ which is the distance between the victim qubits and the target qubits of the source xy-lines. The parameters $m, \Lambda_0$ are from the linear fit to the average of $\Lambda_\mathrm{dir}^\mathrm{(xy)}$ versus $d^\mathrm{(xy)}$.}\label{app_fig_xyqubitdirectcrosstalk}
\end{figure}

Figure \ref{app_fig_xyqubitdirectcrosstalk}(a) shows a histogram of the predicted crosstalk $\Lambda_\mathrm{dir}^\mathrm{(xy)}$ due to direct xy-qubit capacitive interaction. The average value is $-95\,$dB, with the worst at $-49\,$dB, which are much lower than those measured in our processor. This simulation shows that the direct capacitive interaction between a victim qubit and any other source xy-line cannot be the major contributor to the xy-crosstalk level measured in our processors, at least within the simplified model used here.

We have plotted the distance-dependent xy crosstalk (due to the direct xy-qubit interaction) in fig.\,\ref{app_fig_xyqubitdirectcrosstalk}(b), similarly to fig.\,\ref{fig4} in the main text. The data exhibits a relatively large spread of $\Lambda_\mathrm{dir}^\mathrm{(xy)}$ vs.\@ $d_{i,j}$ (distance between the victim qubit $i$ and the target qubit of the source xy-line $j$). We performed an empirical linear fit of the average crosstalk $\bar{\Lambda}_\mathrm{dir}^\mathrm{(xy)}$ vs.\@ $d_{i,j}$ and obtained the distance-scaling parameters $m_\mathrm{xy}=-8.8\,$dB/mm and $\Lambda_0=-57\,$dB, well below the values needed to bring the total single-qubit gate error below the $0.1\,\%$ threshold for 100-qubit processors ($m_\mathrm{xy}=-2.0\,$dB/mm and $\Lambda_0=-50\,$dB; see Section \ref{main_section_distancescaling} of the main text, and Appendix \ref{app_1qerror} below).

As discussed previously, the model used in this simulation does not include the wirebonds and the PCB signal traces. While they are positioned further away from the qubits, they are also much bigger than the rest of the signal lines in the processor. Including them in the model is computationally expensive, and future investigations should look into the strength of the interaction between the combined wirebonds and PCB system with the qubits.

\section{Crosstalk between signal lines}\label{app_linecrosstalk}
\begin{figure}
\centering\includegraphics{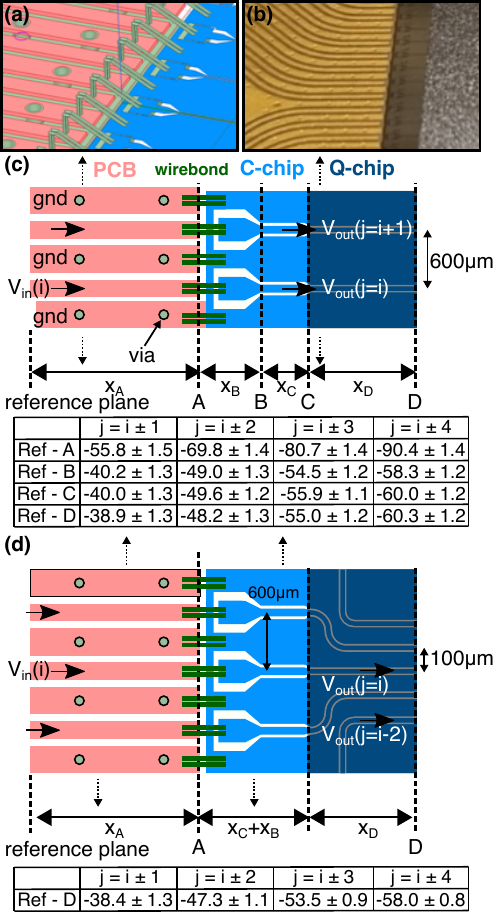}
  \caption{Simulating crosstalk between signal lines. (a) The actual simulation model showing a part of the PCB (with its vias), the wirebonds, the C-chip with its signal lines. The Q-chip is not shown, but is included in the model when required. Five signal lines are included in the simulation. (b) Photograph of the actual packaging and the chips (wirebonds are not shown). (c) A simplified model of (a) showing an input port [$V_\mathrm{in}(i)$] and two output ports [$V_\mathrm{in}(j)$]. This drawing is not to scale. The table summarises the simulated S-parameter $S(j,i)$ (in dB) between the input port $i$ at the reference plane and the output port $j$ at one of the locations (A, B, C, and D). The model assumes $x_A=5.0$\,mm, $x_B=500\,\upmu$m, $x_C=650\,\upmu$m, and $x_D=2$\,mm. (d) A model that is similar to (c) but with the signal lines within the flip-chip environment spaced by $100\,\upmu$m to replicate the actual separation between lines in our processors. Here $x_A=5.0$\,mm, $x_B=500\,\upmu$m, $x_C=650\,\upmu$m, and $x_D=5$\,mm. This drawing is not to scale.}\label{app_fig_linecrosstalk}
\end{figure}

In this section, we are interested in the expected rf-crosstalk level between neighbouring signal lines. To do so, we performed electromagnetic simulation of a small-scale model shown in fig.\,\ref{app_fig_linecrosstalk}(a). The model includes a representative section of the PCB signal traces, wirebonds, and a section of the chips. No tunnel structure is included in this simulation.

Figure \ref{app_fig_linecrosstalk}(c) shows a simplified version of the model. We model five neighbouring signal lines with input ports located at the PCB side (labelled `reference plane' or `ref') and consider signal output ports at various locations. The latter is to illustrate the relative contribution from various parts of the signal lines. The quoted crosstalk levels $S(j,i)=V_\mathrm{out}(j)/V_\mathrm{in}(i)$, expressed in dB, are the average between the highest and lowest values obtained for the considered pairs, and the uncertainty is half of that range. The various locations correspond to the boundary between the PCB and the C-chip (location A, no wirebonds and chips in the model), after the launchpad (location B, with wirebonds and C-chip, no Q-chip in the model), just before the Q-chip (location C, but no Q-chip is included in the model), and within the flip chip environment (location D). The simulations employ ANSYS HFSS\,\cite{zotero-4428} (driven modal, maximum $\Delta S:0.005$, $4$\,GHz frequency, and the outer boundary of the model set by default to the \textit{Perfect E} boundary condition). Due to the limited computational resources, we only include a certain section of the PCB ($x_A=5.0\,$mm) and the Q-chip ($x_D=2.0$). When simulating the results for intermediate locations, e.g. location A, we removed the rest of the signals lines (from B all the way to D) from the model.

First of all, we see that up to the edge of the PCB (location A, row `Ref-A' in fig.\,\ref{app_fig_linecrosstalk}(c)), nearest-neighbour line crosstalk ($j=i\pm1$) is at the level of $-56$\,dB with rapidly decreasing crosstalk down to the level of $-90\,$dB for third next-nearest neighbour lines ($j=i\pm4$). However, after including the wirebonds and the launchpad area (row `Ref-B'), the crosstalk level worsened substantially to the level of  $-40\,$dB for nearest neighbours and $-58\,$dB for third next-nearest neighbours. Subsequent addition of signal lines at the chip level (`Ref-C', `Ref-D') do not substantially worsen the crosstalk even though we have only simulated a $2\,$mm portion of the signal lines within the flip-chip environment. The relatively small change in the crosstalk level for the transmission lines within the chip can be explained by the much narrower transmission lines used at the chip area ($\sim10\,\upmu$m excluding the launchpad structure) compared to the line-to-line separation that is $600\,\upmu$m.

In fig.\,\ref{app_fig_linecrosstalk}(d), we modified the model for the neighbouring signal lines within the flip-chip environment to be spaced by $100\,\upmu$m corresponding the actual separation in our processor layout. To account for the additional space required to bend the signal lines and bring them closer to each other, we extended $x_D$ to $5\,$mm. We also note that $x_D$ is close to half the width of the Q-chip ($12\,\upmu$m). The simulation results show a crosstalk level that is not substantially worse than those obtained in  fig.\,\ref{app_fig_linecrosstalk}(c).

In our processor design, there is no pair of qubits controlled by xy-lines that are nearest neighbours as we always need at least a signal line for flux control of the tunable coupler in between two qubits. For any pair of nearest-neighbour qubits, they are controlled by xy-lines that are \textit{at least next-nearest neighbours}. Typically the corresponding qubits are located along the same row. In the index notation introduced in fig.\,\ref{app_fig_linecrosstalk}, they correspond to pairs of ${i,j}$ with $j=i\pm2$. For next-nearest qubits, they are controlled by xy-lines that are at least in third next-nearest neighbour configuration, i.e. $j=i\pm4$. 

The average xy-crosstalk level measured for pairs of qubits that are controlled by next-nearest neighbour and third next-nearest neighbour xy-lines are $-36.2\,$dB $-38.5\,$dB respectively. Simulation results from fig.\,\ref{app_fig_linecrosstalk}(d) ($-47\,$dB for next-nearest, $-58\,$dB for third-nearest) suggest that the crosstalk between signal lines is not the dominant source of crosstalk, at least within the simplified model considered here.

The accuracy of this result is limited by the model in which we consider only five signal lines on a small section of the combined PCB and QPU system. In addition, the actual layout of neighbouring signal lines is not quite as simple as illustrated in fig.\,\ref{app_fig_linecrosstalk}(c,d). However, these simplified models are already very challenging to simulate given the available computational resources and the large range of physical sizes that need to be taken into account (from few tens of $\upmu$m up to few tens of mm). It is possible that the overall crosstalk level may change when we consider an area that is larger than the one considered here. It is also unclear if the used convergence criteria are adequate to yield results that are representative of the actual processor. In the future, it will be useful to test the accuracy of such simulations on smaller-scale devices before employing it to predict performance of much bigger devices. It will be also interesting to look for alternative modelling techniques that can yield sufficiently reliable results with a much lower computational cost.

\section{Empirical linear fit of $\bar{\Lambda}^\mathrm{(xy)}$ vs.\@ $d^\mathrm{(xy)}$}\label{app_empiricalfit}

\begin{figure}[t!]
\centering\includegraphics{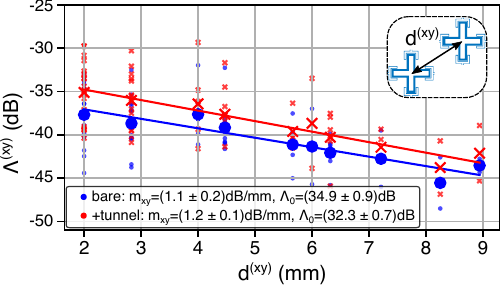}
  \caption{Data of xy crosstalk $\Lambda^\mathrm{(xy)}$ vs.\@  qubit-qubit distance $d^\mathrm{(xy)}$ with an empirical, log-linear fit (i.e.\@ empirical here means that the fit expression is not based on a physical model or numerical simulation of $\Lambda^\mathrm{(xy)}$).}\label{app_fig_empiricalfit}
\end{figure}

To further illustrate the trend within the data of average xy crosstalk $\bar{\Lambda}^\mathrm{(xy)}$ versus the qubit-qubit distance $d^\mathrm{(xy)}$, we performed an empirical linear fit of the form $\bar{\Lambda}^\mathrm{(xy)}=m_\mathrm{xy} d^\mathrm{(xy)}+\Lambda_0$. The fit is performed by assigning equal weight to each data point. The results are shown in fig.\,\ref{app_fig_empiricalfit}. The associated error bars of $m_\mathrm{xy}$ and $\Lambda_0$ are the fit uncertainties. The analyses yield approximately similar $m_\mathrm{xy}$ and $\Lambda_0$ for both processors.

\section{Single-qubit gate error}\label{app_1qerror}

We introduced the distance-scaling parameters, denoted as $m_\mathrm{xy}$ and $\Lambda_0$, which are obtained from an empirical linear fit to the data $\bar{\Lambda}^\mathrm{(xy)}$ vs.\@ $d^\mathrm{(xy)}$. These parameters can be used as a starting point in estimating the impact of xy crosstalk on the total gate fidelity during parallel applications of single-qubit gates. From the hardware-architecture perspective, these parameters should be as low as possible. We find approximately $m_\mathrm{xy}=-1.1\,$dB/mm and $\Lambda_\mathrm{0}=-33.9\,$dB. 

\begin{figure}
\centering
\includegraphics{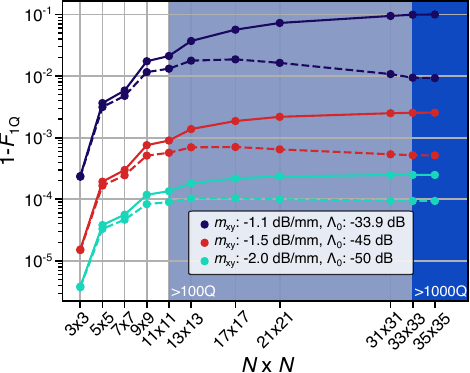}
  \caption{Predicted total single-qubit gate error in a square lattice of $N\times N$ qubits (nearest-neighbour separation is 2\,mm). The gate error is assumed to only come from the off-resonant excitation due to adversary single-qubit signals applied on other xy-lines. The solid line represents a scenario described in the text, and the dashed line assumes additional phase related to propagation from the location of the source line to the victim qubit.} \label{app_fig_1qerror}
\end{figure}

Clearly, the actual error depends on the pulse sequences (implementing quantum gates) that are being applied to all of the qubits. Here, we consider a specific scenario to obtain a concrete estimate of the effect on gate fidelity: a situation where single-qubit $X$ gates are simultaneously applied on all qubits. The qubit frequencies are allocated according to the two-frequency subgroup strategy shown in fig.\,\ref{fig2}(f) of the main text and specified in Appendix \ref{app_targetfrequencies}. Here, the gate error is due to parasitic excitation (both off-resonant and on-resonant) of the victim qubit from adversary $X$-gates applied to all other source xy-lines. These signals are assumed to be on-resonant with the qubits that they are meant to excite. Details on the simulation are described in Appendix \ref{app_numericalsimulations}. 

The results are shown in fig.\,\ref{app_fig_1qerror}. The simulation yielding the solid lines assumes no additional phase offset between the signal and the adversary signals. In this case, the total gate error begins to plateau (up to considered processor sizes) as we begin to consider processors with hundreds of qubits, which is attributed to weakening crosstalk due to larger values of $d^\mathrm{(xy)}$. This assumption of no phase offset is not a realistic one, but we believe it provides an easily understandable setting and a relatively stringent criterion to give us an idea of the total gate error. 
If we could improve the crosstalk and its scaling behaviour to $m_\mathrm{xy} = -1.5\,\mathrm{dB/mm}$ and $\Lambda_0 = -45\,\mathrm{dB}$, for processors with roughly 1000 qubits, we find a projected total single-qubit gate fidelity in excess of 99.7\,\%, and for  even better crosstalk, $m_\mathrm{xy} = -2.0\,\mathrm{dB/mm}$ and $\Lambda_0 =-50\,\mathrm{dB}$, we find 99.97\,\%.

We also simulated the total single-qubit gate error (dashed lines in fig.\,\ref{app_fig_1qerror}) by assuming additional phase offset due to signal propagation between the positions of the source xy-line (approximated as the position of the target qubit of the source xy-line) and the victim qubit. 
The assumption makes a negligible difference for small qubit arrays due to the small $d^\mathrm{(xy)}$. 
The difference is more pronounced for larger qubit arrays, as there are more contributions from victim-source pairs with larger $d^\mathrm{(xy)}$. 
In such a setting, the total gate fidelities for processors beyond 1000 qubits ($33\times33$) increase to 99.95\,\% for $m_\mathrm{xy} = -1.5$\,dB/mm, $\Lambda_0 = -45$\,dB and 99.99\,\% for $m_\mathrm{xy} = -2.0$\,dB/mm, $\Lambda_0 = -50$\,dB. 

We emphasize that this is a very simplified model; we have no quantitative model predicting a log-linear behaviour nor that a similar trend would continue beyond the range of $d_\mathrm{xy}$ examined in this work. However, we believe that these parameters can be useful as a set of quantitative metrics to help guide near-term scaling-up effort and to guide hardware roadmaps.

\section{Numerical simulations}\label{app_numericalsimulations}
In this section, we briefly describe the models used to numerically simulate the ac-flux calibration curve and the total single-qubit gate error due to xy crosstalk.

\subsection{ac-flux calibration curve}
In this simulation, we focus on a system comprising a fixed-frequency qubit coupled to a frequency-tunable coupler. The goal is to extract the Ramsey-fringe frequency of a qubit when the coupler is being subjected to specific dc-flux bias and ac-flux modulation.

The system Hamiltonian $\mathcal{H}_\mathrm{sys}$ is
\begin{align}
    \mathcal{H}_\mathrm{sys} = \mathcal{H}_\mathrm{q} + \mathcal{H}_\mathrm{c} + \mathcal{H}_\mathrm{int},
\end{align}
where $\mathcal{H}_\mathrm{q}$, $\mathcal{H}_\mathrm{c}$, and $\mathcal{H}_\mathrm{int}$ represent the Hamiltonian of the qubit, the coupler, and their interaction, respectively. We employ a doubly rotating frame at the qubit-drive angular frequency $\omega^\mathrm{(xy)}$ and the coupler angular frequency at its dc-flux bias, $\omega_\mathrm{c}(\Phi=\Phi^\mathrm{(dc)})$. 

The qubit Hamiltonian is the following
\begin{align}
    \mathcal{H}_\mathrm{q}/\hbar = -\Delta_\mathrm{q}(\sigma_\mathrm{z}\otimes \mathbb{I}) + \Omega^\mathrm{(xy)}(t)(\sigma_\mathrm{x}\otimes \mathbb{I}),
\end{align}
where $\omega_\mathrm{q}$ is the qubit angular frequency, $\Delta_\mathrm{q} = \omega^\mathrm{(xy)}-\omega_\mathrm{q}$,  $\Omega^\mathrm{(xy)}(t)$ is the time-dependent qubit-drive signal appropriately expressed in the angular-frequency unit, and $\sigma_\mathrm{x/z}$ are the Pauli matrices.

The coupler Hamiltonian is
\begin{align}
    \mathcal{H}_\mathrm{c}/\hbar = -\Delta_\mathrm{c}(t)(\mathbb{I}\otimes\sigma_z),
\end{align}
where $\Delta_\mathrm{c}(t)=\omega_\mathrm{c}(\Phi^\mathrm{(dc)}) - \omega_\mathrm{c}(\Phi^\mathrm{(dc)}+\Phi^\mathrm{(ac)}(t))$, and $\omega_\mathrm{c}(\Phi)=\omega_\mathrm{c0}\sqrt{|\cos{(\pi\Phi/\Phi_0)}|}$. The parameters $\omega_\mathrm{c0}$ and $\Phi_0$ are the coupler angular-frequency at zero-flux bias and the magnetic-flux quantum,  respectively. 

The interaction Hamiltonian is
\begin{align}
    \mathcal{H}_\mathrm{int}/\hbar = ge^{i(\omega^\mathrm{(xy)}-\omega_\mathrm{c}(\Phi^\mathrm{(dc)}))t}(\sigma^+\otimes\sigma^-) + \mathrm{H.c.},
\end{align}
where $g$ is the coupler-qubit coupling strength expressed in the angular-frequency unit. Here, we make the rotating-wave approximation to neglect the counter-rotating terms.

We simulate the state of the combined system (both qubit and coupler are initially in the ground state) evolving under $\mathcal{H}_\mathrm{sys}$ with the pulse sequence shown in fig.\,\ref{app_fig_acfluxcrosstalk}(a). The solver is \textit{qutip.sesolve} \cite{johansson2012, johansson2013}, and the simulation time step is 1\,ns. From the expectation value of $\sigma_z\otimes\mathbb{I}$ at the end of the pulse sequence, we reconstruct the expected Ramsey fringes. An example of the simulation result is shown in fig.\,\ref{app_fig_acfluxcrosstalk}(c).

\subsection{Single-qubit error}
In this simulation, consider a system comprising a square array of qubits (total $N\times N$ qubits), each of them being driven by an on-resonant $X$ gate. Due to non-zero xy crosstalk, each qubit also experiences parasitic qubit driving from the rest of the qubits. 

We assume distance-dependent xy-crosstalk behaviour that follows the empirical linear model described in Appendix \ref{app_empiricalfit}. We focus on a victim qubit at the center of this array (assume $N$ is an odd number). Our goal is to calculate the resultant single-qubit error on this victim qubit due to the parasitic drives.

The Hamiltonian of interest considers a single qubit that is being driven by microwave signals. We ignore evolution of other qubits, and group the parasitic drives into the drive term. Specifically,
\begin{align}
    \mathcal{H}_\mathrm{sys} = \mathcal{H}_{i} + \mathcal{H}_{i,j}.
\end{align}
We employ a reference frame rotating with the drive frequency of the victim qubit.

The Hamiltonian $\mathcal{H}_{i}$ is
\begin{align}
    \mathcal{H}_{i}/\hbar = -\Delta_{i}\sigma_\mathrm{z} + \Omega_{i,i}(t)\sigma_x,
\end{align}
where $\Delta_{i}= \omega_{i}^\mathrm{(xy)}-\omega_{i}$ is the qubit-drive detuning, assumed to be zero in this simulation. The parameter $\Omega_{i,i}(t)$ is the time-dependent microwave drive pulse amplitude applied to the victim qubit by its own xy-line and is appropriately expressed in the angular-frequency unit. The symbols $\sigma_\mathrm{x/y/z}$ are the Pauli matrices.

The Hamiltonian corresponding to the parasitic drives from other qubits is
\begin{align}
    \mathcal{H}_{i,j}/\hbar = \sum_{j\ne i}\Omega_{i,j}(t) (\sigma_x\cos{(\phi_{j}(t))+\sigma_y\sin{(\phi_{j}(t))}},
\end{align}
where $\phi_{j}(t)=\delta_{i,j}t+\phi_{i,j}$. The parameters $\delta_{i,j}$ and $\phi_{i,j}$ are  the detuning ($\omega_j-\omega_i$) and the additional phase, respectively. For simplicity, each qubit drive is assumed to produce the same Rabi angular frequency to its own qubit. Thus $\Omega_{i,i}(t)=\Omega^\mathrm{(xy)}(t)$ for any $i$. The parasitic Rabi amplitude $\Omega_{i,j}$ is
\begin{align}
    \Omega_{i,j}(t)=10^{\Lambda_{i,j}/20}\Omega^\mathrm{(xy)}(t),
\end{align}
where $\Lambda_{i,j}=m_\mathrm{xy}d_{i,j}+\Lambda_0$, and $d_{i,j}$ is the distance between q$_i$ and q$_j$. The phase $\phi_{i,j}$ is the phase related to crosstalk between the source xy-line xy$_j$ and the victim qubit q$_i$. In fig.\,\ref{app_fig_1qerror}, we simulated two specific cases: $\phi_{i,j}=0$ (solid line), and $\phi_{i,j}=(\omega_j/c)d_{i,j}$ (dashed line) corresponding to the phase delay due to signal propagation from the location of the target qubit of the source xy-line to the location of the victim qubit, through vacuum. 

We employ \textit{qutip.sesolve} solver \cite{johansson2012, johansson2013}, with a simulation time step of 0.125\,ns. From the expectation value of $\sigma_\mathrm{z}$, we extract the probability $P_\mathrm{g}$ of the qubit being in the ground state at the end of the pulse sequence. We repeat this simulation for a victim qubit with the eight different frequencies in Appendix \ref{app_targetfrequencies}, and we average the probabilities. This is the single-qubit gate error ($1-F_\mathrm{1Q}$) that is plotted on the vertical axis of fig.\,\ref{app_fig_1qerror}.

\bibliography{xtalk_paper}

\end{document}